\pdfoutput=1 
\documentclass[
floatfix,
superscriptaddress,
 aps,
reprint
]{revtex4-1}

\usepackage{graphicx}
\usepackage{dcolumn}
\usepackage{bm}
\usepackage{amsmath}
\usepackage{color}
\usepackage{wasysym}
\usepackage{float}
\usepackage{array,url,kantlipsum}
\usepackage{verbatim}
\usepackage{xcolor}
\usepackage{CJK}
\usepackage{textcomp}
\newcolumntype{P}[1]{>{\centering\arraybackslash}p{#1}}
\usepackage{array}
\usepackage[normalem]{ulem}

\newcolumntype{x}[1]{>{\centering\arraybackslash\hspace{0pt}}p{#1}}

\begin{document}

\title{Sensitive magnetometry in challenging environments}
\author{Kai-Mei C. Fu}
\affiliation{University of Washington, Physics Department and Electrical and Computer Engineering Department, Seattle, WA, 98105, USA}
\author{Geoffrey Z. Iwata}
 \affiliation{Helmholtz-Institut, GSI Helmholtzzentrum f{\"u}r Schwerionenforschung, 55128 Mainz, Germany}
\affiliation{Johannes Gutenberg-Universit{\"a}t Mainz, 55128 Mainz, Germany}
\author{Arne Wickenbrock}
 \affiliation{Helmholtz-Institut, GSI Helmholtzzentrum f{\"u}r Schwerionenforschung, 55128 Mainz, Germany}
\affiliation{Johannes Gutenberg-Universit{\"a}t Mainz, 55128 Mainz, Germany}
\author{Dmitry Budker}
 \affiliation{Helmholtz-Institut, GSI Helmholtzzentrum f{\"u}r Schwerionenforschung, 55128 Mainz, Germany}
\affiliation{Johannes Gutenberg-Universit{\"a}t Mainz, 55128 Mainz, Germany}
\affiliation{Department of Physics, University of California, Berkeley, CA 94720-7300, USA}

 \date{\today}

\begin{abstract}

State-of-the-art magnetic field measurements performed in shielded environments with carefully controlled conditions rarely reflect the realities of those applications envisioned in the introductions of peer-reviewed publications. Nevertheless, significant advances in magnetometer sensitivity have been accompanied by serious attempts to bring these magnetometers into the challenging working environments in which they are often required. This review discusses the ways in which various (predominantly optically-pumped) magnetometer technologies have been adapted for use in a wide range of noisy and physically demanding environments.

\end{abstract}

\maketitle

\section{Introduction}
Magnetic fields are routinely measured with high sensitivity to probe the physical processes that underlie a vast array of natural phenomena. From geological movements, solar flares, and atmospheric discharge, to inter-cellular processes, neuronal communication in the brain, molecular scale chemical processes, and sub-atomic interactions; magnetic fields are produced and can convey detailed information about both the dynamic and static properties of these systems.

While the ubiquity and high information density of these fields make them attractive to measure, many of the field sources to be studied exist in environments that are hostile to the state-of-the-art precision magnetometry techniques developed over the last 60 years. Magnetometers with sensitivities better than 10\,pT$/\sqrt{\mathrm{Hz}}$ must contend with measuring signals in environments with formidable magnetic noise and dramatic gradients. Whether in a laboratory, hospital, airport, moving car, helicopter or spacecraft, the background environmental noise must be understood and surmounted. Neurons firing in the brain have an associated magnetic field signal on the scalp surface lasting a couple ms with a few hundred fT amplitude (i.e., only parts per billion of the Earth field), meaning movement within even a 1\,nT/cm gradient would swamp the monitored brain signal.

Beyond these difficulties of signal-to-noise, magnetometers may also be required to perform under physical conditions which at first glance may seem incompatible with their operating principles. High- and low-pressure and temperature environments affect the performance of almost any magnetometer. High levels of radiation, whether in space, at particle accelerators or at nuclear reactors, could disrupt the sensitive quantum states prepared in magnetometers, or compromise sensor infrastructure and electronics. Vibrations can make certain measurements impossible, while a rapidly changing field may be far beyond a sensor's bandwidth and dynamic range. We must also consider the compatibility of the sensor with the environment itself - it would be a difficult task indeed to convince a doctor to place a hot rubidium or cesium vapor cell inside a patient's body, no matter how well protected.

Unsurprisingly, utilization of enhanced measurement techniques and noise mitigation is an integral part of magnetometer development across many applications, and is well documented across the literature. Notwithstanding, recent advances in sensors motivate new criteria for evaluating the best match between sensor and application, and the techniques to optimize measurements. While no single magnetometer type is suited for all sensing challenges, the diversity in available magnetometer technologies can cover the envisioned plethora of environments. 

Here, we describe how optically pumped magnetometers (OPMs) based on hot atomic vapors as well as negatively charged nitrogen-vacancy (NV$^-$) defects in diamond are particularly well suited for these challenges, and the strategies that have been developed to adapt these sensors to the broad range of measurement conditions. 

\section{Optically-pumped magnetometers}
\label{sec:op_mags}
\begin{figure*}[t]
\centering
\includegraphics[width=\textwidth]{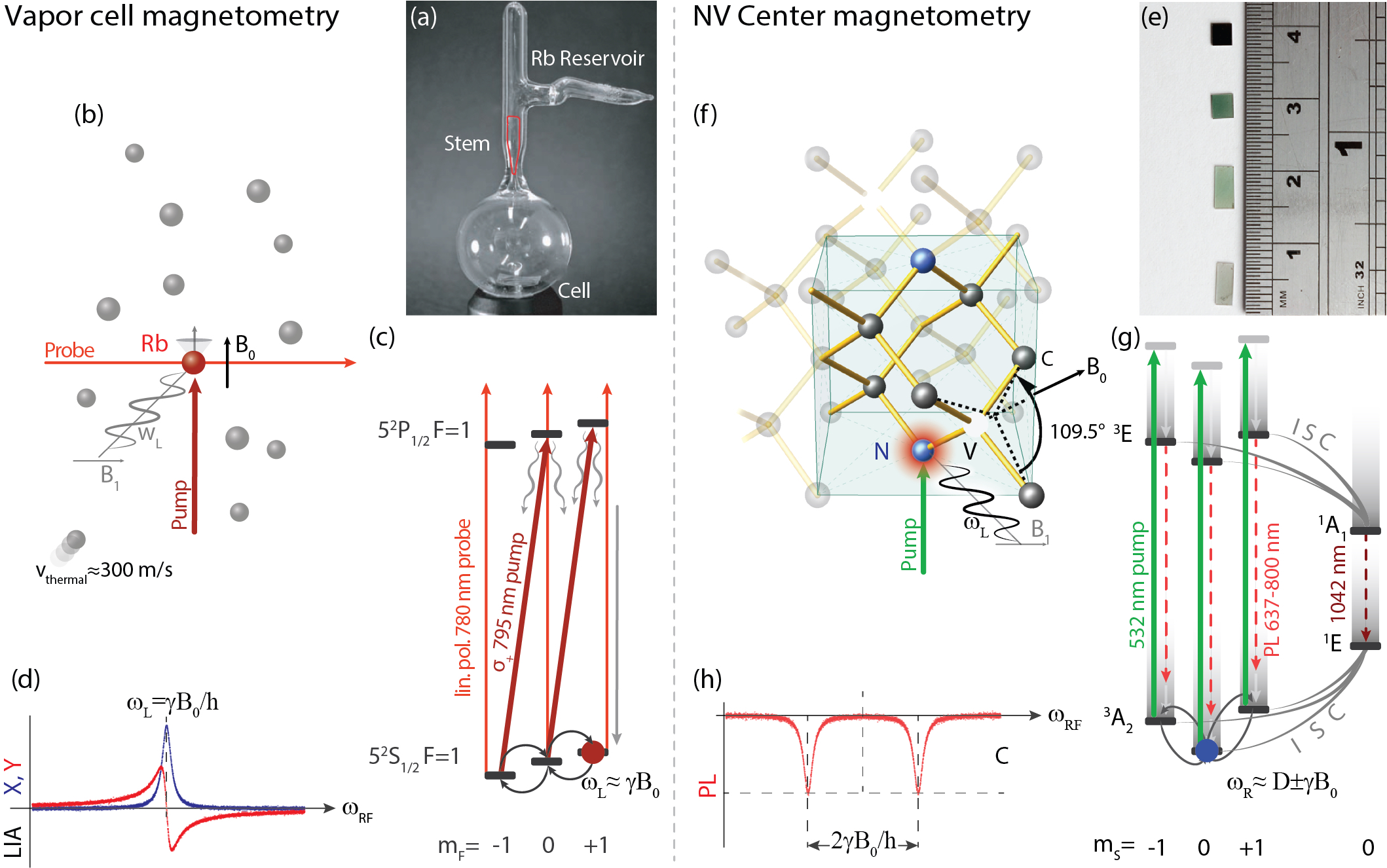}
\caption{Optical magnetometry with vapor cells (a-d) and with NV centers (e-h) emphasising the similarities. (a) An example of a glass vapor cell used for magnetometry. It is two centimeters in diameter, has a reservoir for Rb and a stem to separate the sensing volume and the reservoir. The inner walls of the cell are coated with an alkene film \cite{Balabas2010Minute} enabling coherence times of up to 77\,s. (b) Basic working principle (admitting numerous variations). Individual Rb atoms flying with thermal velocities ballistically through the cell are subject to pump laser light to polarize the magnetic moments, a magnetic field $B_0$, an oscillating magnetic $B_1$-field to probe the magnetic resonance and an additional probe laser field to read out the spin-precession non-destructively via optical rotation of its polarization. (c) The Rb magnetic-sublevel manifolds of the $F=1$ ground and excited states. Optical pumping with a circular polarized on-resonant D1-line laser light at 795\,nm transfers the atoms into the $m_F=+1$ sublevel (quantization axis along $B_0$). If the oscillating $B_1$-field corresponds to the energy difference between Zeeman sublevels, there is a coherent transverse component of the magnetization that manifests itself as polarization oscillation at the Larmor frequency $\omega_L$ in the far detuned linear polarized probe beam. (d) The corresponding magnetic resonance appears after demodulating this oscillating signal with a lock-in amplifier (LIA). Its center frequency is, to first approximation, proportional to the magnetic field. This magnetic-field measurement scheme is referred to as M$_z$ magnetometry\,\cite{Budker2013ombook}.\\
(e) A selection of diamond samples having been exposed to different doses of electron irradiation but not yet annealed. Irradiation and annealing are steps in producing samples with different color-center densities. Room-temperature coherence times can be milliseconds. (f) Basic working principle (admitting numerous variations). NV centers are embedded in a diamond matrix and conventionally subjected to green 532\,nm light for polarization, a magnetic field $B_0$, and an oscillating magnetic $B_1$-field to probe the magnetic resonance. The measured signal from the NV centers is either spin-state dependent red photoluminescence (PL) or infrared absorption (not shown). (g) The relevant energy levels of the NV centers: the magnetic triplet ground and excited states ($^3A_2$ and $^3E$, respectively) separated by an energy corresponding to the 637\,nm zero-phonon line and the two singlet levels $^1E$ and $^1A_1$. The grey gradients above the energy levels indicate broadening due to phonons. A green 532\,nm laser is often used to excite the $^3A_2\rightarrow^3E$ transitions. Even though the excitation and subsequent PL are spin-conserving, the intersystem crossing (ISC) to the singlet states is not. The rates are higher for the $m_S=\pm 1$ excited-state sublevels, which leads to accumulation of NV population in the $m_S=0$ sublevel. Magnetic resonances can be observed as a reduction of photoluminescence when the frequency of the oscillating magnetic field $B_1$ matches the energy difference between magnetic sublevels. On resonance, population is transferred from the $m_S=0$ into one of the $m_S=\pm1$ sublevels. The $m_S=\pm1$ states have a higher energy due to spin-spin interactions. (h) The two spin transitions can be observed around a central frequency of $D\approx 2870$\,MHz, with the difference of the two center frequencies determined by the magnetic field. Important quantities are the contrast $C$ and the width of the resonance as a measure for the coherence time of the NV center (ensemble).}
\label{Fig:workingprinciple}
\end{figure*}
There is a large variety in the ways to measure magnetic fields in extreme environments. In this review, we discuss mainly (but not exclusively) two types of devices that belong to the same broad class of optically pumped magnetometers (OPM): magnetometers based on glass cells filled with vapors of alkali atoms and magnetometers utilizing color-center defects in solids and more specifically, the workhorse of the field, nitrogen-vacancy (NV) centers in diamond. Both magnetometer types implement precise measurements of the response of the respective spin system to magnetic fields, to infer the local magnetic field. Both magnetometer types can be initialized (``pumped'') to a specific spin state, controlled and read out with laser light (see Fig.\,\ref{Fig:workingprinciple}). Both types of devices have received growing attention in recent years and are being applied for measurements in various challenging environments due to their unique properties. 

Vapor-cell magnetometry, with a demonstrated sensitivity of $\approx160\,\text{aT}/\sqrt{\text{Hz}}$ (10$^{-18}\,\text{T}/\sqrt{\text{Hz}}$) by M.\,Romalis' group at Princeton~\cite{Romalis2010}, is comparable to state-of-the-art SQUID systems (superconducting quantum interference devices) which require cryogenics \cite{Drung2017,Schmelz2016_SQUID}. The sensitivity of vapor-cell magnetometers improves with size; for a given density the more atoms one can probe the better. This scaling is advantageous for spatially extended field measurements with the trade-off of limited spatial resolution. These sensors usually incorporate glass cells filled with gas, which are macroscopic objects with dimensions typically between 0.2 and 10\,cm (Fig.\,\ref{Fig:MagnetomDifferentSpatialScales}c) and are not well suited for probing magnetic structures below these dimensions.

NV-center magnetometers are typically less sensitive than vapor cells, although continuously improving~\cite{barry2020son}. This lower-sensitivity is offset by the flexibility provided by the solid-state host matrix. The solid-state sensors operate from cryogenic temperatures~\cite{schaefer-nolte2014dbs} to over 700\,K~\cite{toyli2013mcs} and from high-vacuum~\cite{schaefer-nolte2014dbs} to 60 GPa~\cite{doherty2014epm}. Moreover, there are sensors based on single NV centers nanometers from the diamond surface~\cite{mamin2013nanoscale}, in nanofabricated scanning probe tips~\cite{maletinsky2012rsd,Degen2008}, and in diamond nanoparticles (nanodiamond)~\cite{zhang2018hnq}. This flexibility enables unprecedented small sample-sensor distance, nanometer spatial resolution and incorporation into biological systems. While the solid-state focus of this review is the NV-center in diamond, we note that the field is under rapid development. Solid-state magnetometers have also been demonstrated in SiC, a technologically advanced wide-band gap semiconductor~\cite{son2020dsc}, and electrical readout of magnetic sensors in both diamond and SiC has been realized~\cite{simo2016aod,kraus2014mft,christle2015ies,niethammer2016vmu}. 

In the following sections, we discuss how OPMs can be used in various challenging environments including high-noise environments, physically extreme environments and biological environments. 

Magnetic sensors employ elements such as heated vapor cells or cooled cryogenic components,  as well as microwave and low-frequency magnetic fields. They can thus influence a nearby sample or other proximal sensors, which could be a problem for dense sensor arrays required for applications like magnetoencephalography (Sec\,\ref{subsec:brain_magnetometry}) or for experiments searching for parity- and time-reversal-invariance-violating electric dipole moments \cite{Abel2019OPM}. Radiofrequency fields used in magnetometers can also exert unwanted influence on the sample and adjacent sensors. This problem can be mitigated with the methods discussed in the following subsections.

\subsection{All-optical magnetometry}
\label{subsec:all-optical}

All-optical magnetometry, where no other fields but light are applied to the sensing element, in the case of atomic-vapor magnetometers goes back to the seminal work of Bell and Bloom \cite{Bell_Bloom1961} who realized that, for atoms, the effect of microwave fields can be replaced by that of modulated light. Various kinds of modulation were studied and applied to magnetometers over the years \cite{Alexandrov2005Dynamic,Grujic2015FSPmag}. An example of the necessity to use all-optical techniques is stand-off and mesospheric magnetometry (Sec.\,\ref{subsubsection:space_and_remote}).   
It turns out that, in certain cases, not only scalar but also all-optical vector magnetometry is possible \cite{Patton2014all,Zhivun2014RFmag}. Here the idea is that modulated bias magnetic fields are replaced with effective magnetic fields produced by light shifts.

\subsection{Microwave-free diamond magnetometry}
\label{subsec:MW-free}

\begin{figure}[t]
\centering
\includegraphics[width=\columnwidth]{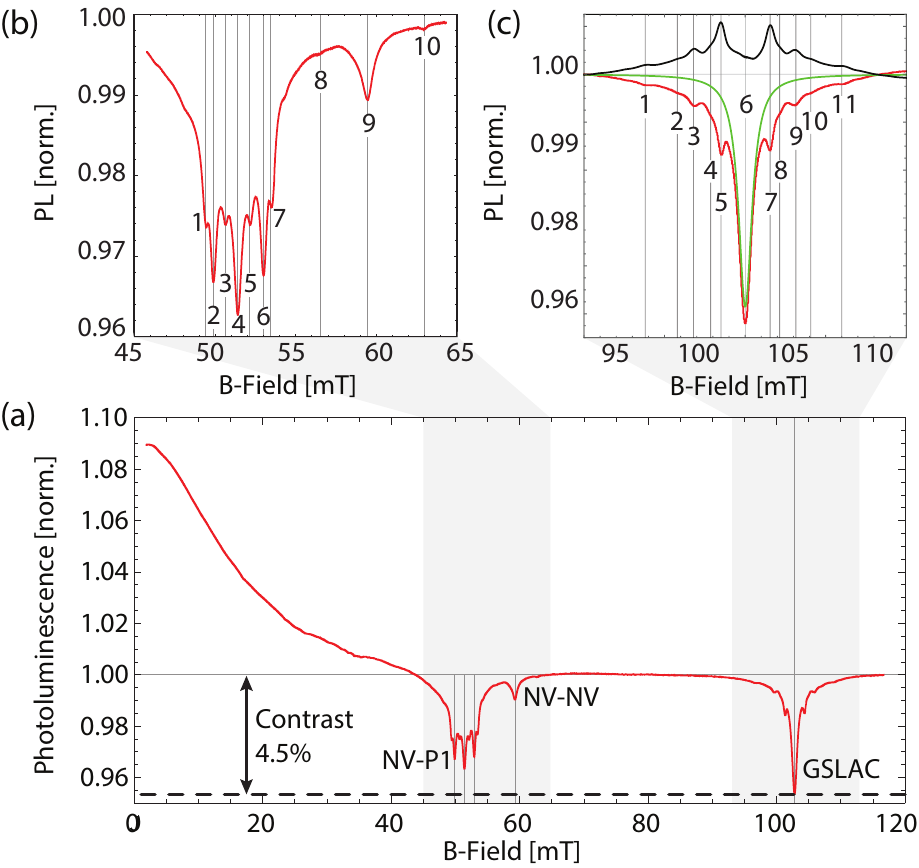}
\caption{Photoluminescence of a an ensemble of NV centers illuminated with green light as a function of applied magnetic field. (a) An ``overview'' scan over a broad magnetic field range. (b) Detail around 51.2\,mT. Visible are features that can be attributed to cross-relaxation with substitutional nitrogen atoms, so called P1-centers (1-7). Features 8-10 are due to cross-relaxation with NV centers oriented along other three axis of the diamond crystal. (c) Detail around the ground-state level anticrossing (GSLAC) where the Zeeman-shifted electronic $m_S=-1$ state is close in energy to the $m_S=0$ state \cite{Zheng2017LevelCrossing,Auzinsh2019GSLAC}. (Reprinted from \cite{wickenbrock2016microwave} , with permission of AIP Publishing.)}
\label{Fig:NVfluorvsB}
\end{figure}
All-optical magnetometry has so far been possible with NV centers either with relatively low sensitivity using the broad dependence of NV fluorescence near zero field (see Fig.\,\ref{Fig:NVfluorvsB}(a) or at cryogenic temperatures, where NV centers exhibit the effect of electromagnetically induced transparency (EIT) (see Ref.\,\cite{Acosta2013EIT} and references therein). At higher temperatures, optical transitions of NV centers preserve spin projections and no EIT effects have been observed so far.
 Magnetometry can  still be performed without applying microwave fields employing a bias dc magnetic field to ``tune'' near one of the sharp features seen in Fig.\,\ref{Fig:NVfluorvsB}. The sharp dependence of the signal on magnetic field near these features enables magnetometric sensitivities comparable to those obtained with the techniques employing microwave fields  \cite{wickenbrock2016microwave,Akhmedzhanov2017MWfree}. Extensions of the MW-free magnetometry technique include its combination with cavity-enhanced absorption readout \cite{Zheng2017LevelCrossing} and MW-free \textit{vector} magnetometry \cite{Zheng2020MWfreeVector}.
We note in passing that a technique for microwave-free \textit{electric} sensing with NV centers was proposed recently \cite{Block2020optically}. 

\section{The signal-to-noise challenge}
\label{sec:s_to_N_challenge}

Magnetometers with high sensitivity are attractive for measuring minute signals, but an experimenter must also contend with the environment in which the measurement will be made.
Relevant magnetic noise sources are ubiquitous in the laboratory, field, and urban environment. Power-distribution lines produce oscillating magnetic fields with amplitudes 10-20\,nT at a distance of 30 meters. A car passing by, or an elevator moving at a distance of 10 meters produces fields of a few nT. Even away from man-made sources of noise, one needs to contend with natural noise of the geomagnetic field which is typically at the 1\,pT/$\sqrt{\text{Hz}}$ level at 1\,Hz (with approximately $1/f$ increase towards lower frequencies) as well as gradients on the order of 20\,pT/m or even larger \cite{harrison1991magnetic}. This necessarily creates challenges for measuring signals at the fT level.
In addition to multitudes of transient noise sources, real-world environments may also contain background fields that are large in magnitude or have large gradients. These features can create serious technical challenges for magnetometers with limited dynamic range, can be additional sources of noise and can compromise magnetometer performance.

The general approach to addressing the ``signal-to-noise'' challenge consists in a combination of measures to enhance the signal and reduce  the noise.
In this section we review the common techniques utilized in the magnetometry community to isolate sub-nT level magnetic signals from the noisy environment. Many techniques are engineering tools common to all magnetic-sensor platforms, however for concreteness we include examples in specific platforms in which each technique has been successfully implemented. We conclude the section with a discussion of emerging hybrid sensors in which  correlating different sensor modalities may be particularly useful when dealing with challenging environments.

\subsection{Flux engineering}
\label{subsection:Flux_engineering}

Flux engineering provides a suite of signal enhancement tools specific to magnetic sensing and includes flux concentrators, flux guides and flux transformers. A flux concentrator utilizes a high-permeability material to compress the magnetic flux from a larger area into a smaller sensor area. The technique is successfully used in a variety of sensors including Hall probes~\cite{leroy2007umc}, magnetoresistance sensors~\cite{edelstein2006ptt,caruso1998npm}, magnetic tunnel junctions~\cite{chaves2008mob}, superconducting quantum interferometer devices (SQUIDs)~\cite{bondarenko2002fcm}, atomic-vapor magnetometers~\cite{griffith2009mam} and, most recently, NV-based magnetometers~\cite{fescenko2019dme}. 

\begin{figure*}[t]
\centering
\includegraphics[width=\textwidth]{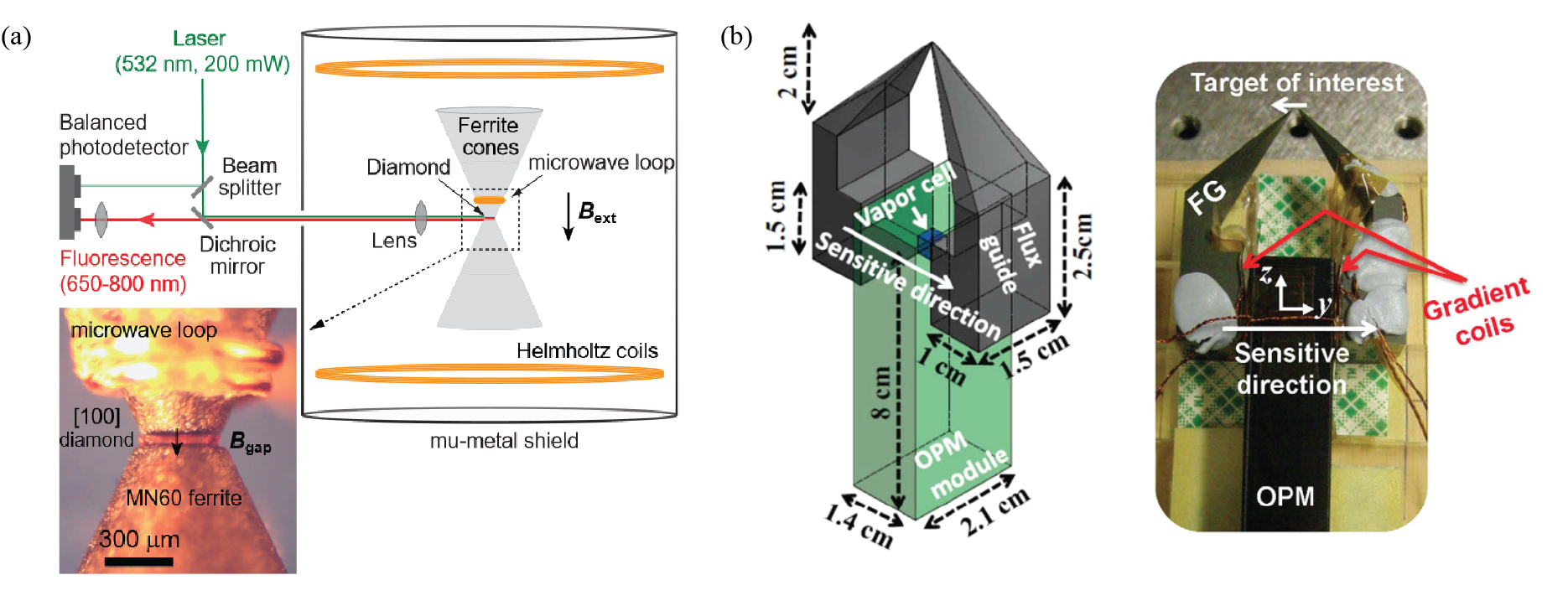}
\caption{Flux concentrators used in conjunction with optically pumped magnetometers: in (a) to enhance the sensitivity to spatially homogeneous magnetic fields in NV-center magnetometry \cite{fescenko2019dme} and in (b) to increase the spatial resolution of a vapor-cell magnetometer \cite{Kim2016}.}
\label{fig:flux}
\end{figure*}

The amplification factor $\epsilon$ obtained utilizing flux-concentration techniques can approach three orders of magnitude~\cite{leroy2007umc, pannetier-lecoeur20078rfr}, however there are practical trade-offs. The sensing area is fundamentally increased, limiting the sensor's spatial resolution. For example, for the flux-concentrator NV magnetometer system (Fig.\ref{fig:flux}a), the flux concentrator had dimensions $\approx$30 times or more of those of the diamond sensor~\cite{fescenko2019dme}. The concentrator-enhanced sensitivity could be, in principle, alternatively achieved by scaling up the sensor size, but  this approach may be practically limited by the available light power and the field-homogeneity requirement. Additional factors that should be taken into account when using magnetic flux concentrators include the filtering of off-axis fields, nonlinearity and saturation, remnant fields and sensitivity to temperature via ferrite thermal noise. A detailed analysis taking into account thermal eddy current and magnetization noise showed that for low frequencies and low-conductivity ferrite materials, magnetization noise dominates~\cite{fescenko2019dme}. Magnetization noise scales linearly as the square root of the magnetic loss factor $\left(\mu''/\mu'^2\right)$ in which $\mu = \mu' - i \mu''$ is the complex magnetic permeability. Most recently, 195 fT/$\sqrt{\text{Hz}}$ sensitivity has been achieved with a flux concentrator utilizing a low-noise permalloy and a 472 magnification factor~\cite{xie2020hmt}. Further significant improvements will require the identification of lower-noise materials.  
Despite these disadvantages, there are several reasons flux concentrators may be desirable in practical devices. Most field applications do not require fT sensitivity and the use of a concentrator relaxes power, size, and weight requirements. We note that a second novel application of flux concentrators is signal modulation (Sec.~\ref{subsection:mod_and_filtering}). The signal is modulated via the magnetic permeability by utilizing a small coil wrapped around the ferrite concentrator~\cite{qasimi2004imh} or mechanically moving the concentrator with a micro-electromechanical systems (MEMS) device~\cite{edelstein2006tst}.
Related concepts to that of flux concentrators are those of flux guides \cite{Kim2016} (Fig. \ref{fig:flux}b) and flux transformers \cite{Savukov2014mft}. A flux guide enables the use of a larger atomic OPM with high sensitivity for a smaller sensor area. This solves the issue of reduced sensitivity of small cells. A flux transformer can be used to decouple the ac fields to be measured from dc or slowly varying fields and gradients in the measurement region that can adversely affect the performance of the sensor. This is a particularly useful ``trick''  in the case of magnetic resonance imaging (MRI), where fields and gradients are used to polarize the nuclear spins and encode information on their spatial location~\cite{Savukov2014mft}.

\subsection{Magnetic shielding}
\label{subsec:Shielding}

Ideally, to eliminate noise, one could shield all environmental magnetic fields from the sensor. Metals with high relative magnetic permeability (mu-metal) can be shaped and layered to form shielding which can effectively block out magnetic fields at all frequencies, creating an ideal environment for magnetometers to focus on a signal of interest. For human-size experiments, the state of the art is the BMSR-2, an eight-layer magnetically shielded room at the Physikalisch-Technische Bundesanstalt (PTB) in Berlin with a passive shielding factor of 10$^6$ at frequencies $>$0.01\,Hz and with a typical residual magnetic field $<1$\,nT~\cite{thiel2007dms}. 
One critical component for performance is degaussing the shielding layers. With an improved degaussing system with distributed coils, residual fields as low as $<100$\,pT are theoretically possible with only a three-layer room and residual fields $<$130\,pT over a (0.5\,m)$^3$ cubic enclosure have been realized~\cite{sun2020llm}. Nanotesla-level shielding is typically utilized in fundamental physics experiments (e.g. the search for electric dipole moments of atoms, molecules, and the neutron) and for biomagnetic measurements such as neural imaging utilizing magnetoencephalography~\cite{Yashchuk2013shielding,sun2020llm,hari2012mfs}. 

\begin{figure}[t]
\centering
\includegraphics[width=\columnwidth]{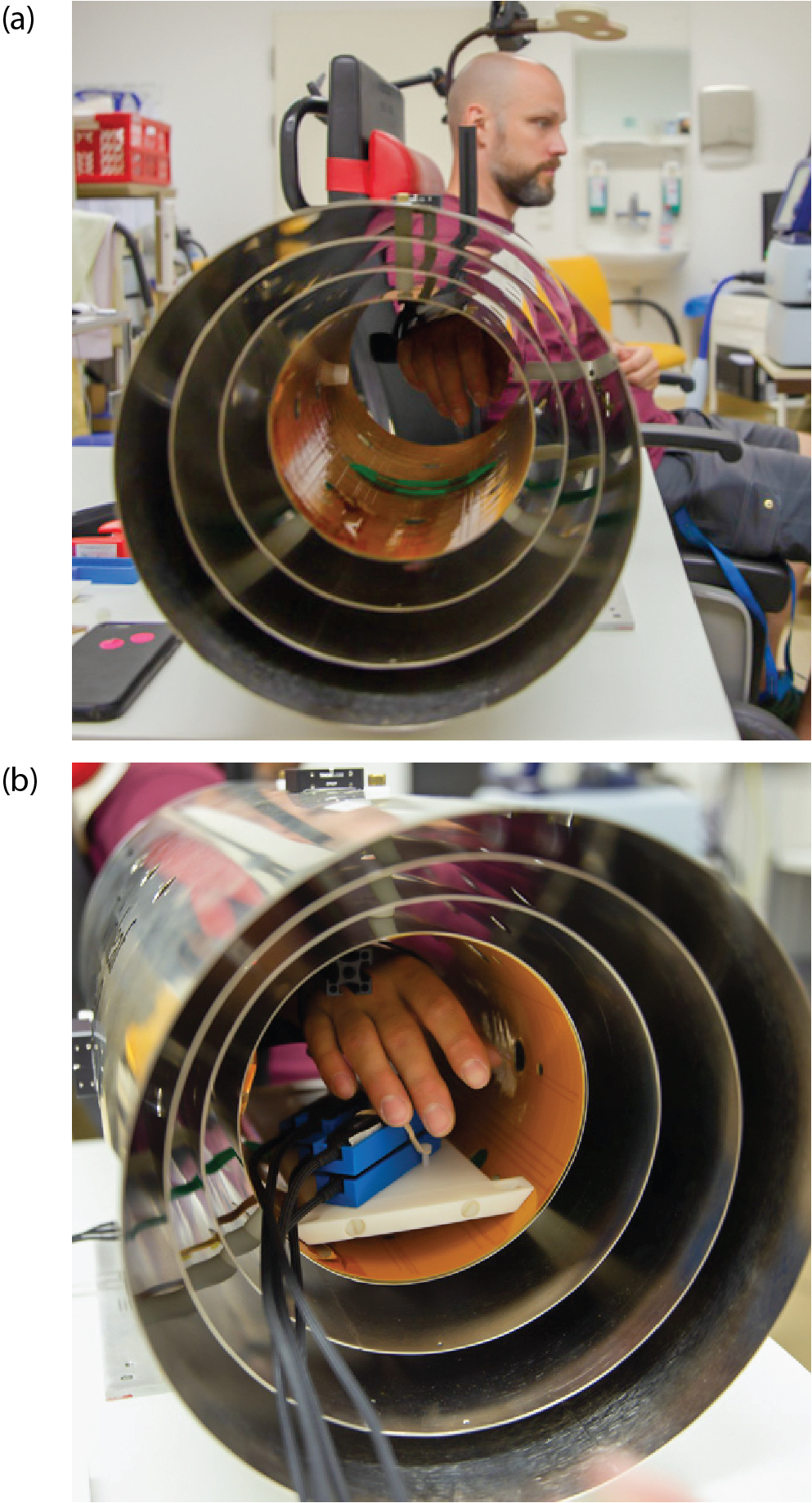}
\caption{(a) Biomagnetic measurements in a hospital environment with partial magnetic shielding. The recorded magnetic signals in the subject's hand inserted in a shield are triggered by transcranial magnetic stimulation. (b) Several commercial atomic zero-field OPMs (in blue casing) are used to measure the spatial distribution of biomagnetic signals from a stimulated muscle.}
\label{fig:arm}
\end{figure}

As magnetometers themselves become smaller and more portable, so should the shielding solutions. Transcranial magnetic stimulation (TMS)-evoked muscle activity has recently been detected inside a hospital setting~\cite{iwata2019bsr} utilizing a portable shield which only encompasses the forearm and hand of the subject~\cite{broser2018opm} (Fig.~\ref{fig:arm}). This application highlights the utility of small, compact sensors and shields. TMS requires a high-intensity magnetic pulse applied to the brain which must be shielded for the detection of 10\,pT level signals at the hand. 

\subsection{Signal modulation and filtering}
\label{subsection:mod_and_filtering}
Filtering is a simple solution to remove time-varying noise in a frequency band that has minimal overlap with the signal band. Filtering can be utilized to remove both magnetic noise or non-magnetic noise which presents as a magnetic signal. As discussed above, examples of environmental transient magnetic noise sources include power lines, trains, elevators, communication signals, as well as noise of the geomagnetic field. Noise sources are also typically present in the sensor apparatus; for example, the stability of the applied bias field, the optical excitation intensity, and RF control amplitude and phase. Standard techniques to filter this noise is to include notch filtering, adaptive notch filtering~\cite{keshtkaran2014fra,widrow1975anc}, independent component analysis~\cite{ikeda2000ica}, variation of the interrogation time relative to the noise period, reference measurements on timescales shorter than the noise correlation time ({\it e.g.} used in Ref.~\cite{wolf2015sdm}), and ``phase cycling'' as most frequently discussed in the nuclear magnetic resonance (NMR) context \cite{Tseytlin2017_phase_cycling}. There are, however, clear limitations to filtering. Knowing the nature of the noise does not always allow one to simply ``filter it out". In measuring biomagnetic transients, for example, a 50\,Hz (60\,Hz in the USA) notch filter over an action potential lasting 10\,ms would cause significant distortion of the physiological signal, since any band limitation in the frequency domain creates artifacts such as ringing, phase lag, or signal attenuation in the time domain \cite{parks1987digital,widmann2012filter}. 

\begin{figure}[h!]
\centering
\includegraphics[width=\columnwidth]{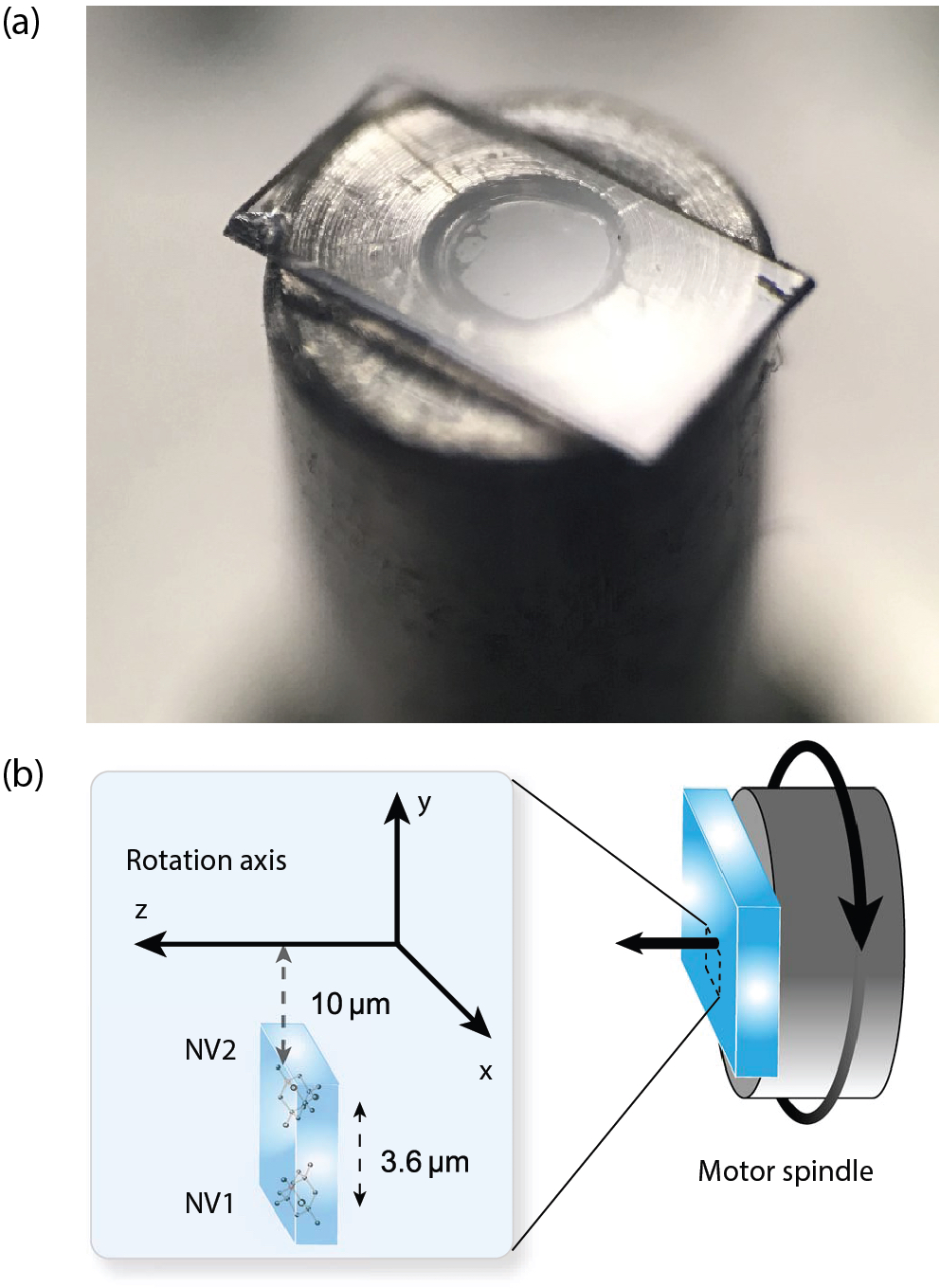}
\caption{Diamond-sensor rotation rates exceeding 3000\,kHz have enabled $T_2$-limited sensing of a dc field~\cite{wood2018t2l} and rotation sensing~\cite{wood2020oqp}.  (a) Diamond on a drill (Image credit A. Wood / U. Melbourne) (b) Measuring rotation with a diamond. From Ref.\,\cite{wood2018qmr}}.
\label{fig:rotation}
\end{figure}

Filtering is often combined with signal modulation, which shifts the signal from a noisy low-frequency band to a higher-frequency band to enable the use of notch filtering or lock-in detection method. Signal modulation techniques relevant for OPM devices are modulations of the optical excitation~\cite{Alexandrov2005Dynamic} (amplitude, frequency and polarization),  RF frequency~\cite{fescenko2019dme},  magnetic field~\cite{failache2003iei}, and flux amplification~\cite{qasimi2004imh} (Sec.\;\ref{subsection:Flux_engineering}). Mechanical rotation of the entire diamond sensor is another potential modulation technique~\cite{wood2018qmr} (Fig.~\ref{fig:rotation}). In some cases, such as materials characterization \cite{ROMALIS2011_Materials}, the sample itself can  be rotated or vibrated in order to move the signal into frequencies where $1/f$ noise is reduced. Indeed, frequency modulation is often referred to as frequency ``upconversion'' since the low-frequency signal is converted to a higher frequency. In one novel upconversion method, the NV-center in diamond is used to detect a static transverse magnetic field via the detection of the oscillating field due to the Larmor precession of a nearby nuclear spin~\cite{liu2019nvd}.

One potential advantage of detecting upconverted ac signal is that one can utilize rephasing techniques to extend the spin coherence and thus improve the sensitivity of the sensor.  For ac magnetic sensing, the sensitivity is limited by $T_2> T_2^*$. Here, $1/T_2$ and $1/T_2^*$ are the homogeneous and inhomogeneous relaxation rates of the spins, respectively. The ac magnetic sensitivity is further enhanced with dynamical decoupling pulse sequences, pioneered by the NMR community, used to rephase the magnetometer spins~\cite{viola1999ddo}. Alternatively, this rephasing technique can be described in the language of filtering~\cite{biercuk2011dds} or quantum lock-in amplification~\cite{kotler2011siq}. Enhancements up $T_2/T_2^*$ could be naively expected, however often the same noise fields that limit $T_2^*$ will also be upconverted~\cite{liu2019nvd}. The advantage of enhanced interrogation time however remains. While spin-rephasing techniques are commonly used in solid state magnetometry, recently, the use has been extended to alkali atoms in solid parahydrogen~\cite{upadhyay2020usc}.

\subsection{Common-mode-noise rejection}
\label{subsection_Common-mode-noise_rejection}

Common-mode noise can be filtered utilizing more than one measurement. This strategy exploits a dependence on magnetic field between the two measurements which differs either in sign or magnitude from other noise sources. Here we discuss some specific examples of this strategy: differential magnetometry, gradiometry, networks of sensors, double-quantum and comagnetometry.

\subsubsection{Differential magnetometry, gradiometry, sensor networks}
\label{subsubsection:Differential_magnetometry}

Differential measurement is perhaps the most common approach to dealing with noisy environments \cite{Veryaskin2018}. 
Suppose, for instance, one needs to make a measurement of a local source (e.g. human heart or brain). The magnetic fields produced by local sources fall rapidly with distance $r$ from the source (as $r^{-3}$ for a source that can be approximated as a point dipole or $r^{-2}$ for a linear dipole), while unwanted fields from distant sources possess a much larger degree of ``spatial coherence.'' If we have a sensor next to the source of interest, ideally, at a distance no greater that the spatial extent of the source, then the signal on the proximal sensor is a sum of the signal of interest produced by the source and a signal from spurious distant sources. Because of the larger spatial extent of the latter, that part of the signal can be subtracted using the signal from a second, reference sensor at a distance much larger than the spatial extent of the source. In the case where the field of interest and the background both vary in time sufficiently slowly, the second sensor may not even be needed if it is possible to \textit{move} a single sensor to alternate the source plus background and background measurements.

\begin{figure}[H]
\centering
\includegraphics[width=80mm]{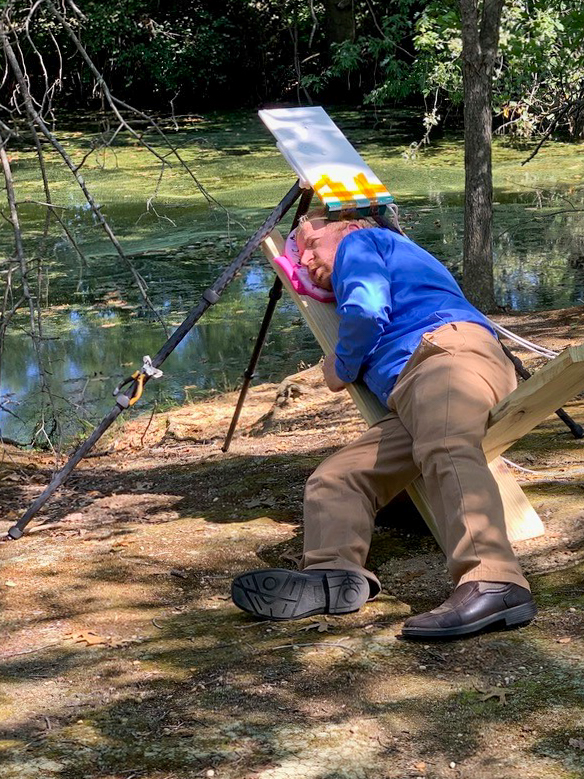}
\caption{Field test of an OPM designed for recording magnetic signals from the brain \cite{Limes2020portable}. Professor M. V. Romalis of Princeton University was a volunteer subject for the test. Photo courtesy Dr.\,Tom Kornack (Twinleaf, twinleaf.com).}
\label{fig:opmMEGfield}
\end{figure}

A logical continuation of this approach is a magnetic gradiometer--a device that is insensitive to a uniform field, but rather measures the spatial variation of the field. A ``synthetic'' gradiometer can be constructed from separate individual sensors, and, in some cases, it is possible to construct a sensor that is intrinsically a gradiometer. An example of a modern intrinsic gradiometer used for brain measurements in an unshielded environment (see Sec.\,\ref{sec:bio}) is that described in Ref.~\cite{Zhang2020pig}. The authors constructed a scalar gradiometer using two miniature  (5$\times5\times5$\,mm$^3$ inner volume) cesium vapor cells separated by 5\,cm, and vertical-cavity surface-emitting lasers (VCSELs) producing independent pump and probe beams for both cells. Off-resonant linearly polarized probe light interrogated the two cells at the same time, so that the output of the gradiometer was proportional to the change of the magnetic field over the baseline given by the distance between the cells. The use of the same probe beam to interrogate the two cells has a benefit of rejecting common-mode noise related to the probe laser beam (this could be useful also for non-gradiometer measurements, as recently demonstrated, for example, with a so-called light-shift dispersed $Mz$ sensor \cite{Oelsner2020integrated}). The reported gradiometric sensitivity of this device was better than $18\,$fT/cm/$\sqrt{\text{Hz}}$, sufficient for reliably detecting biomagnetic signals from a human body in unshielded environment. In fact, OPM gradiometers have already been used to detect human brain signals outdoors \cite{Limes2020portable}, as shown in Fig.\,\ref{fig:opmMEGfield}. A single-cell intrinsic magnetic gradiometer with a sensitivity of $10\,$fT/cm/$\sqrt{\text{Hz}}$ was reported in \cite{Lucivero2020femtotesla}.

With multiple sensors, one can set up a high-order gradiometer, and, as a further extension of the idea, an array of sensors (for instance, as a wearable brain senor, see Ref.~\cite{Tierney2019} and references therein) and even a magnetometer network. Such networks are capable of detecting signals with complicated spatio-temporal signatures and find applications in geophysics \cite{Angelopoulos2008}, fundamental physics \cite{Afach2018,MASIAROIG2020}, and even in the study of magnetic signatures of cities (Sec.\,\ref{subsection:urban_magnetometry}). 
It should be pointed out that extraction of maximal amount of useful information from a network of sensors is generally a highly nontrivial task (see, for example, Ref.~\cite{MASIAROIG2020} discussing the problem in the context of the Global Network of Optical Magnetometers for Exotic Physics Searches (GNOME); see also Sec.\,\ref{subsection:urban_magnetometry}) and benefits from modern data-analysis techniques like machine learning. 

\subsubsection{Double-quantum magnetometry}
\label{subsubsection_Double-quantum magnetometry}

Double-quantum (DQ) magnetometry utilizes a pair of spin transitions whose frequency response is identical to most non-magnetic perturbations but differs for magnetic signals. For the specific case of NV centers in diamond, these transitions correspond to the ``single-quantum'' $m_s = 0 \leftrightarrow m_s=-1$ and $m_s = 0 \leftrightarrow m_s=+1$ transitions (Fig.\;\ref{Fig:workingprinciple}(g)). The two transitions have identical dependence on electric field, strain, pressure, and temperature, while an equal but opposite dependence on magnetic field (and rotations about the NV axis) \cite{dolde2011efs,mittiga2018ilc,VMA2010,broadway2019mis}. Thus by subtracting the two single-quantum signals, fluctuations in the non-magnetic field are eliminated while magnetic sensitivity is enhanced (see, for example, \cite{Fescenko2019_Hemozoin} in which this technique is used wide-field magnetic imaging of hemozoin crystals). In a similar vein, one can add the two signals to obtain enhancement of a non-magnetic signal (for example, for thermometry~\cite{SingleNVThermometry, toyli2013fte, neumann:2013hpn}). 

\begin{figure}[t]
\centering
\includegraphics[width=\columnwidth]{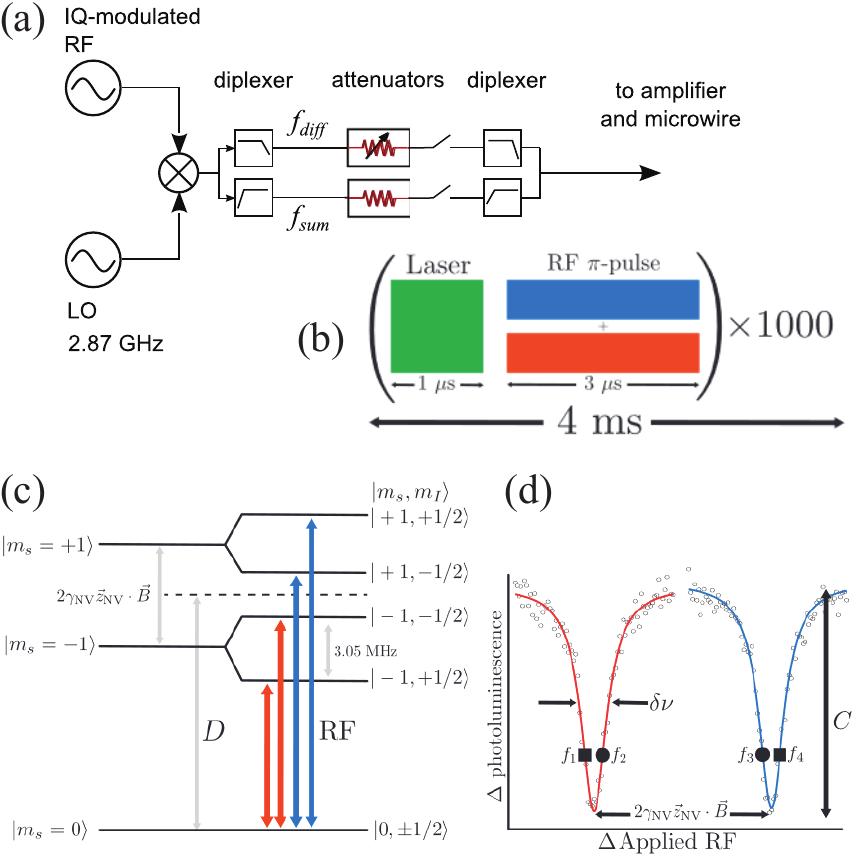}
\caption{(a) The use of dual-frequency pulses in double quantum magnetometry eliminates noise due to temperature and electric field (reproduced from Ref.~\cite{mamin2014mdq}). (b-d) The use of four-frequency pulses ($f_1 to f_4$), biased on the outer and inner edges of the single quantum resonances, enables double quantum magnetometry in imaging applications. (From Ref.~\cite{kazi2020msi}). In this imaging application, a CMOS camera is used with a 4~ms exposure time. The pulse sequence (b) thus runs many times during a single exposure. (c) energy level diagram. (d) bias points in the NV ODMR spectrum.}
\label{fig:Dual_Frequency}
\end{figure}

It is often advantageous to directly create a ``double-quantum'' (DQ) superposition of the $m_s=\pm1$ states. The original double-quantum demonstration utilized a single-tone, broadband RF pulse to create this superposition~\cite{fang2013hsm}. More recent demonstrations utilize two-tone pulses~\cite{mamin2014mdq} (Fig.\,\ref{fig:Dual_Frequency}a) or multiple pulses~\cite{munuera-javaloy2020dqm} that have the advantage of allowing operation at arbitrary bias fields. The DQ technique can be also implemented in a cw mode, in which the two-tone microwaves are modulated, and detection is done with a lock-in method~\cite{fescenko2019dme}. DQ measurements have demonstrated a two-fold enhancement in signal-to-noise~\cite{mamin2014mdq} and a 15-fold improvement in magnetic sensitivity by eliminating thermal drifts~\cite{fang2013hsm} relative to single-quantum magnetometry. The DQ technique was also utilized to eliminate strain gradients and inhomogeneities that present as spurious magnetic signals~\cite{bauch2018udt}. Finally, four-tone RF pulses, Fig.\,\ref{fig:Dual_Frequency}b) were utilized in wide-field imaging applications, in which one frequency is used on each side of the two electron resonances (Figs.\,\ref{fig:Dual_Frequency}\,c,d). This ``double'' DQ magnetometry eliminates the effect of inhomogeneity of both the resonance position as well as the lineshape across the image~\cite{kazi2020msi}.  

\subsubsection{Comagnetometry}
\label{subsubsec:Comagnetometry}

The resonance frequency of a spin is shifted when the sensor undergoes rotation. While this effect is the basis for rotation sensors, it can also be a systematic error for magnetometry, and is especially relevant for moving sensor applications (Sec.\,\ref{sec:movingplatforms}) and spin-based fundamental-physics experiments~\cite{venema1992sce,vasilakis2009lnl,bulatowicz2013lsl,tullney2013csd,hunter2013uep}.  
The ability to distinguish between magnetic field and rotation-induced frequency shifts is thus of significant importance. A solution to this challenge is provided by comagnetometers~\cite{lamoreaux1987nls}. Comagnetometers compare the spin precession frequencies,  $\gamma_{1,2}B + \Omega$, between two species with different, known gyromagnetic ratios $\gamma_{1,2}$. $\Omega$  is the rotation rate.   
Comagnetometry variants include utilizing different atomic species~\cite{limes2018hxc} and different isotopes~\cite{lamoreaux1987nls, alexandrov2004eds,kimball2013dir} and hyperfine levels of the same atomic species~\cite{gomez2020bec,wang2020ssa}. The latter implementation is particularly robust to systematic errors due to magnetic field gradients. When implementing a comagnetometer, a concern is co-locating the two species; otherwise, the measurement can be compromised.  One approach for nuclear spin-based sensors is to use the spins of mixed liquids \cite{Ledbetter2012liquid} or even different spins residing in one and the same molecule \cite{Wu2018comag}.

\subsection{Hybrid modalities}
\label{Subsec:Hybrid}

Challenging measurements may benefit from hybrid modalities, that can both enhance the signal of interest and discriminate noise. We divide hybrid modalities into two categories. 

In the first category are sensors that detect multiple complementary physical signatures of the investigated object or phenomenon. Examples include the combination of wide-field optical microscopy and NV magnetometry \cite{lourette2019nfs,kayci2018qmn} used to realize NV-based  force-induced remnant magnetization spectroscopy (NV-FIRMS) for measuring the magnitude of binding forces between biologically relevant molecules, as well as wide-field NV magnetometry combined in a single setup with a magneto-optical Kerr-effect (MOKE) microscope \cite{lenz2020pts}. The idea of the latter is that MOKE is sensitive to surface magnetization which does not necessarily produce a magnetic field outside of the sample, so combining it with NV magnetometry that measures the field provides a more complete diagnostic picture. 

In the same category is also real-time tracking of a patient's movement using video cameras, enabling magnetoencephalographic recording without sedating the subject \cite{Hill2019QSPIN_Brain}. Eliminating the need for sedation is particularly important when working with children. Comagnetometers using different species (Sec.\,\ref{subsubsec:Comagnetometry}) are, in a sense, also in the same category of hybrid sensors.

In hybrid sensors of the second category, the physical parameter being measured (e.g., position, pressure, temperature) is ``transduced'' into another parameter (e.g., magnetic field, frequency of optical resonance) and measured via sensing the latter.  A striking example of such a hybrid sensor is a nanothermometer composed of NV centers and a magnetic nanoparticle (MNP) \cite{WangPhysRevX2018Hybrid}. This device takes advantage of the ferromagnetic-paramagnetic transition of the nanoparticle material that occurs at a certain temperature. The temperature susceptibility of the NV magnetic resonance in this hybrid device is 14\,MHz/K, an enhancement of over two orders of magnitude over the temperature susceptibility of NV centers themselves \cite{doherty2014temperature}. The demonstrated sensitivity of the hybrid nanothermometer was 11\,mK/$\sqrt{\text{Hz}}$ under ambient conditions. The authors of Ref.~\cite{WangPhysRevX2018Hybrid} envision that hybrid thermometers can be designed to operate from cryogenic temperatures to about 600\,K by tuning the chemical composition of the MNP. 

In another type of NV-magnetic nanoparticle hybrid sensor, the nanodiamond NV magnetometer and MNP are connected by a stimulus-responsive hydrogel spacer (Fig.~\ref{Fig:MagnetomDifferentSpatialScales}f) that changes its length as a function of temperature, pressure or power of hydrogen (pH) of the environment, leading to a change of the magnetic field detected by the NVs \cite{zhang2018hnq}.

Similar ideas can also be used in fundamental-physics experiments. A recent example is Ref.\,\cite{Jiao2020Exotic_Molecular_Rulers}, where the authors search for exotic interactions between two electron spins, that are separated by ``molecular rulers'' of certain lengths, and interrogated via a double electron-electron resonance (DEER) technique. Comparing the DEER signals for different-length separation between the spins allows one to search for any deviations from the usual dipole-dipole interaction between the spins on the nanometer scale.  

\section{Physically challenging environments}

In this section we focus on the physical challenges presented in practical, high-sensitivity magnetometry. One of the major applications of magnetometry is the study of materials~\cite{casola2018pcm, jayarman1983dac}. For these studies, sensors must be able to operate in extreme temperatures and pressures. Size also presents a physical challenge-whether probing a nanoscale phenomenon or astronomical objects. Practical deployment often requires operation on moving platforms requiring sensors robust to vibrations and changes in orientation with respect to the Earth's (relatively) large magnetic field. Finally, space and accelerator applications require sensors robust to radiation. While the focus of this section is the robustness of the sensor's physical magnetic transduction mechanism, we note that the sensor must include control and readout components. Some of the novel methods to eliminate local controls and further enhance robustness to extreme physical environments include all-optical and microwave-free magnetometry which are discussed in Sec.~\ref{subsec:all-optical}, \ref{subsec:MW-free}.

\subsection{Temperature challenges}
\label{subsection:temperature}
Temperature can dramatically change material properties and magnetic phase changes can be observed from cryogenic temperatures to 1000\,K. Additionally, biosensitive measurements can only be performed in a limited temperature range. If a sensor itself has a limited operational temperature range, temperature isolation of the sensor can be utilized to extend the complete sensor package operational range. Examples of this include the use of SQUIDs (requiring cryogenics) in magnetoencephalography~\cite{hari2012mfs} and hot vapor cells for measurement of biological~\cite{Jimenez-Martinez2017mop}, geological~\cite{Romalis2010,ROMALIS2011_Materials} and chemical~\cite{ledbetter2008zfr} samples. Thermal isolation is also an important consideration for keeping the overall power consumption of the device low. However, this isolation comes at the unavoidable cost of reduced spatial resolution and experimental complexity. Thus sensors which function over a wide temperature range are particularly attractive. Optically-pumped magnetometers by definition work far from equilibrium, and for solid-state OPMs this feature enables them to function from cryogenic temperatures all the way up to 700\,K~\cite{toyli2013mcs,liu2019nvd,duan2019dcm}. Vapor-cell magnetometers are indirectly sensitive to temperature as they require a certain and stable alkali vapor pressure (and therefore number density) to perform. Atomic magnetometer in the spin-exchange relaxation free regime even require stable temperatures above 400\,K to perform \cite{savukov2013ser}.  

One clear advantage of NV-based magnetometers is the diamond host. Diamond is stable in air up to 1000\,K~\cite{john2002o1t} and can withstand higher temperature in an inert gas atmosphere~\cite{evans1964std}. The NV center is stable at temperatures exceeding 1350\,K~\cite{chakravarthi2020wnv}. However for all optically-pumped magnetometers, the spin and optical properties change with temperature, limiting the operational range. At room temperature, the NV longitudinal spin relaxation time $T_1$ exceeds 6\,ms and increases at lower temperature~\cite{Jarmola2012_Temp_B_T1}. While the spin-coherence time and thus magnetic sensitivity can approach the $T_1$-limit utilizing dynamical decoupling techniques~\cite{bar-gill2013sse}, dynamical decoupling is only suitable for small-bandwidth ac sensing applications (Sec.~\ref{subsection:mod_and_filtering}). Thus typically at cryogenic and room-temperature, the NV dc sensitivity is limited by noise due to paramagnetic impurities. Room-temperature operation extends NV magnetometry applications to biological systems which are discussed further in Sec.~\ref{sec:bio}. 

The longitudinal relaxation rate of  NV sensors is a  nontrivial  function of the  temperature \cite{Norambuena2018_NV_T1}, including  hitherto unexplained ``plateau'' of nearly constant relaxation rate between a few K and up to 100\,K or higher depending on the density of the color centers in the  sample \cite{Jarmola2012_Temp_B_T1,Mrozek2015_NV_T1}.
At temperatures T$>$500\,K, $T_1$ for NV centers is found to be proportional to T$^{-5.6}$ with $T_1$ ranging from $T_1^{516\,\textrm{K}} = 119\,$\textmu s to $T_1^{944\,\textrm{K}} = 3.8\,$\textmu s~\cite{liu2019cqc}. In practice, at high temperature  magnetic sensitivity is again not limited by $T_1$. Instead, high-temperature sensitivity is limited by non-radiative processes limiting optical-spin readout contrast and also typically restricting NV magnetometer operation to temperatures below 700\,K~\cite{toyli2013mcs, duan2019dcm,plakhotnik2014aot}. The spin coherence, however, is still robust at higher temperatures and single-NV optically detected magnetic resonance (ODMR) using a pulsed heating technique was utilized to measure the phase transition in a single Ni nanoparticle at 615\,K~\cite{liu2019cqc}. 

\subsection{Cryogenic environment}
\label{sec:Cryogenic}

Conventional vapor-cell OPMs cannot operate in the cryogenic environment, though there have been efforts towards magnetometry with paramagnetic atoms in cryogenic solids, e.g. \cite{Kanorsky1996_Cs_solid_He}, liquids, e.g. \cite{Arndt1993_SF}, and gases, e.g. \cite{Hatakeyama2000_Rb_He,Sushkov2008production}.

As discussed above, diamond sensors can operate at low temperatures, and even benefit from prolonged relaxation times \cite{bar-gill2013sse} at cryogenic temperatures. One low-temperature application has been the study of condensed matter phases~\cite{joshi2020qpt, joshi2019mlc, acosta2019ccd, schlussel2018wfi,waxman2014dms}. Another possible application is monitoring fields of superconducting magnets operating at 4\,K. Compact and robust absolute sensors for this application can perhaps be constructed based on the effect of electromagnetically induced transparency in, for example, SiV centers in diamond, enabling all-optical sensing, see Sec.\,\ref{subsec:all-optical}.

Scanning magnetometry (see e.g. Fig.~\ref{Fig:MagnetomDifferentSpatialScales}a) is important for materials research and condensed-matter physics~\cite{hong2013nmN, casola2018pcm}. A 4\,K scanning NV magnetometry experiment \cite{Thiel2016cryoscan}, which allowed quantitative characterization of the nanoscale structure of a vortex in a superconductor, provided a proof of concept for scanning magnetometry in cryogenic environment. This technique was further refined and applied to imaging of magnetization in monolayer (i.e., two-dimensional) van der Waals magnets \cite{Thiel2019science}. Current work aims to extend the temperature range of scanning NV based magnetometry into the millikelvin range \cite{Maletinsky2020Personal}.

\subsection{High pressure}
\label{sec:High_Pressure}
Like temperature, pressure can also dramatically change material properties~\cite{jayarman1983dac}. Magnetometry can elucidate important aspects of these changes in studies of phase transitions and material synthesis~\cite{hazen2018high,wang2014review}, and in the exploration of new superconducting materials. Geophysicists can simulate and study magnetic properties of the Earth's core~\cite{wei2017GC007143}, or understand the formation of different rock types~\cite{dunlop2001rock}, while chemists can study pressure induced changes in the electronic structure of atoms and molecules.

State-of-the-art diamond anvil cells (DACs) can routinely achieve 100s of GPa, and pressures exceeding 1\,TPa have been reported~\cite{Dubrovinskaiae1600341}. 
Most magnetometers cannot withstand high pressures and thus sensing is performed outside of the high pressure environment. Two examples of these {\it ex situ} sensors are SQUID-based and NMR-based magnetometry. In SQUID-based measurements, a superconducting coil is placed around or near the vicinity of the sample  \cite{wang2014review, Girat2010}. While this approach utilizes the excellent magnetic sensitivity of SQUIDs, the cryogenic requirements place constraints on both the possible temperature range of the samples, and the materials of the high-pressure cell \cite{sadykov2008nonmagnetic}. Furthermore, the diameter of the induction loop is much larger than the sample size, and thus does not provide magnetic flux resolution. NMR techniques, which directly probe spin-spin interactions in materials, use an induction coil and gradient bias field~\cite{kitagawa2009spontaneous,lin2015evolution,lee1987nmr}. Induction-coil NMR techniques mitigate the cryogenic constraints and, while not yet utilized in DAC experiments, can achieve sub-mm spatial resolution. However challenges include poor filling factors, sample alignment and the presence of large background fields.

\begin{figure*}[t]
\centering
\includegraphics[width=\textwidth]{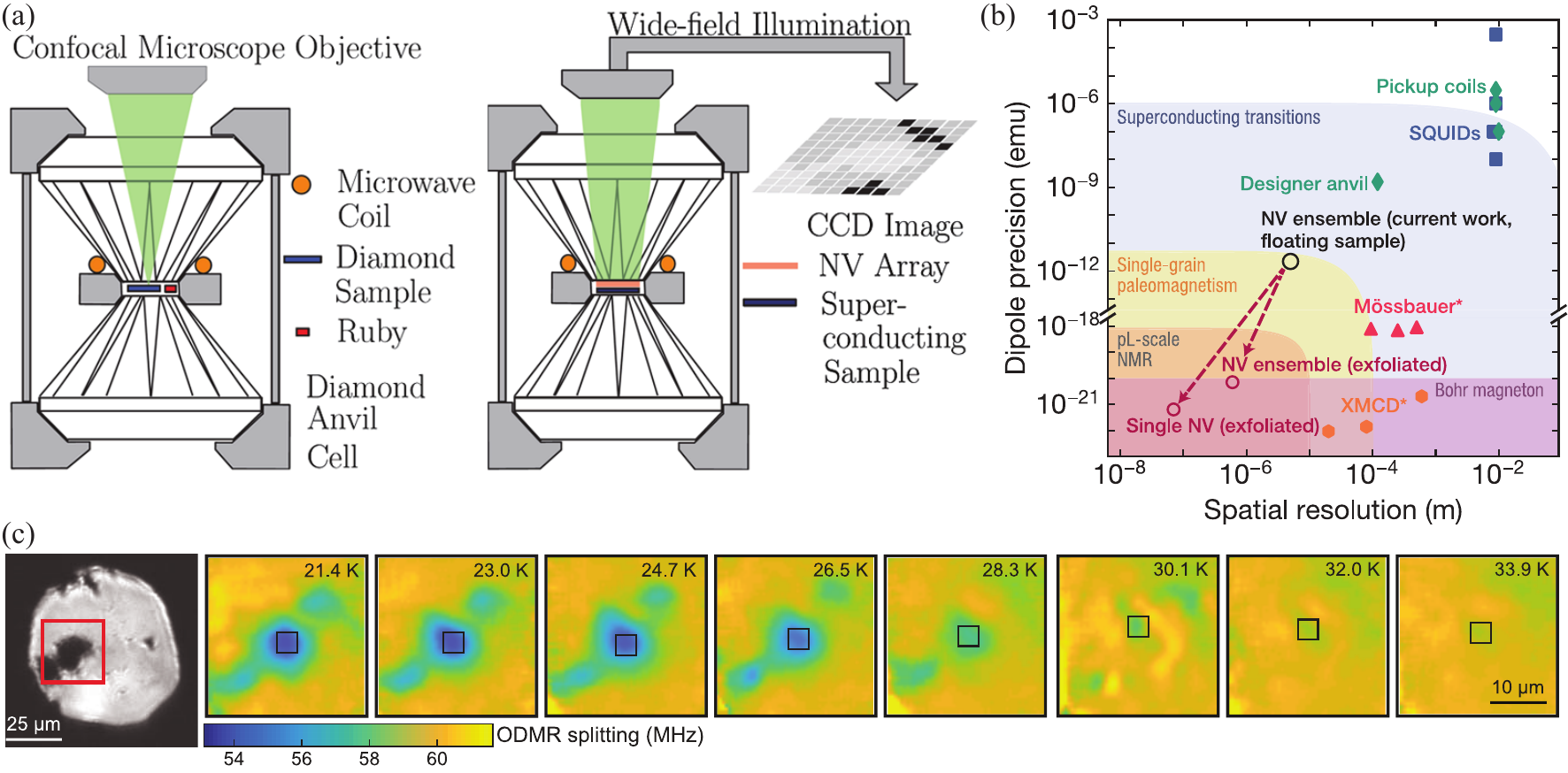}
\caption{(a) Schematic of NV magnetometry measurement in a high-pressure diamond anvil cell from Ref.~\cite{doherty2014epm}. (b) Comparison of high-pressure magnetometry techniques from Ref.~\cite{Hsieh2019} (c) Maps of the ODMR frequency splitting in a diamond sensor above an MgB$_2$ sample at 7\,GPa, reproduced from Ref.~\cite{lesik2019mmm}. Below 30\,K, the exclusion of the magnetic field from the sample is observed.}
\label{fig:High_Pressure}
\end{figure*}

NV spin resonance were observed up to 60\,GPa~\cite{doherty2014epm} and, in contrast to the above techniques, an NV magnetometer in diamond can be placed in close proximity to the sample volume (Fig.\,\ref{fig:High_Pressure}a), even directly integrated into the diamond anvil \cite{lesik2019mmm,Hsieh2019,yip2019mmf}, and spatial resolution can be optically diffraction-limited. A comparison of spatial resolution and magnetic sensitivity for high-pressure magnetometry is given in Fig.\,\ref{fig:High_Pressure}b. Recently integrated NV-DAC systems were utilized to investigate pressure-induced phase transitions in iron~\cite{Hsieh2019,lesik2019mmm} (Fig.~\ref{fig:High_Pressure}c) and Nd$_2$Fe$_{14}$B~\cite{shang2019msi}, the pressure-temperature phase diagram of gadolinium~\cite{Hsieh2019}, and the superconducting transition temperature, critical fields and the Meissner-state local magnetic field profile of BaFe$_2$(As$_{0.59}$P$_{0.41}$)$_2$~\cite{yip2019mmf}. These investigations have focused on high-pressure dc magnetometry; however the extension of NV-based NMR detection~\cite{aslam2017nnm,boss2016otd,devience2015nnm,lovchinsky2016nmr} to DAC systems should enable spatially resolved NMR studies~\cite{steele2017odm}.

\subsection{Low pressure}
\label{sec:Low_Pressure}
Similar to other solid-state magnetometers, and in contrast to atomic vapor magnetometers, low-pressure operation of NV-based magnetometers is straightforward. The aspects of low-pressure magnetometry unique to the NV (or similar defect-based systems) include nanoscale scanning probe imaging~\cite{pelliccione2016spi} and the potential ability to combine low pressure with a large temperature range~\cite{casola2018pcm}.

\subsection{High magnetic fields}
\label{subsec:High_fields}

The techniques to measure high magnetic fields (tens of tesla and higher) were recently reviewed in Ref.\,\cite{battesti2018hmf}, so we only provide a brief summary here.

The strongest DC fields obtained with laboratory magnets are currently around 45\,T \cite{denOuden2016_45T}. With pulsed magnets that are not destroyed during a pulse, one can go to slightly over 100\,T. ``Explosive'' techniques reach to over 2\,kT. Clearly, different techniques are required for measuring magnetic fields in each of these settings.

For DC magnets, NMR teslameters are typically the instruments of choice with the highest relative accuracy of better then a part in $10^{12}$ achieved with magnetometers based on gaseous $^3$He \cite{Nikiel2014}. The absolute accuracy is limited to approximately a part in $10^8$ by the finite accuracy of fundamental constants. The operation range of $^3$He magnetometers is limited to 
about 12\,T due to the effect of ``magnetic decoupling''  of the hyperfine structure (leading to the nuclear spin quantum numbers being conserved in  optical transitions) that reduces the efficiency of optical pumping.

Magnetic fields up to 45\,T are measured with Faraday-induction pick-up coils (during the  ramp-up of the field in the magnet) that are calibrated with NMR ``marker'' lines (see \cite{battesti2018hmf}). NMR marking and Faraday-induction pick-up coils are also used for pulsed 
magnetic fields with typical field-evolution times in the ms to tens of ms range. Calibration can be accomplished using the de Haas - van Alphen effect in copper wherein the magnetization of the material oscillates as a function of magnetic field due to the Landau quantization of electron energy. 

The highest-precision metrology of pulsed magnetic fields is based on the use of rubidium in gas phase \cite{George2017}. The magnetic field ``tunes'' the energies of the Zeeman sublevels in the upper and the lower state of an atomic transition and brings the transition on resonance with light of a fixed and precisely measured frequency, increasing absorption and fluorescence. The instrument of Ref.\,\cite{George2017}, based on the 780\,nm (D2 line) in rubidium, demonstrated an accuracy of $\approx 2 \cdot10^{-4}$ for the entire duration of the pulse.

In the highest-field pulsed magnets \cite{Nakamura2013}, the field is often measured via magneto-optical (Faraday) rotation in a sample of a solid transparent material such as glass.
Faraday-rotation measurements are largely immune to electromagnetic pick-up accompanying the magnetic-field pulse and allows measurement of the field evolution on fast ($\mu$s) scales with large dynamic range. 

\subsection{Near-zero magnetic fields}
\label{subsec:Near-zero_Field}

Certain applications including searches for exotic interactions (e.g., those due to electric-dipole or gravitational-dipole moments of particles, see \cite{Safronova2018}), biomagnetic measurements in shielded rooms (Sec.\,\ref{sec:bio}), or detection of zero- to ultralow-field (ZULF) NMR \cite{Blanchard2016emagres} require operating magnetic sensors, including gradiometers \cite{Jiang2019Grad}, at near-zero field. Indeed, some sensor types, for example, the most sensitive alkali vapor based spin-exchange relaxation-free (SERF) OPM (see, for example, \cite{savukov2013ser}) typically \textit{require} operation at fields below 10$^{-7}$\,T. On the other hand, it was originally thought that diamond NV center based sensors lose first-order sensitivity to magnetic fields at fields below $\approx 0.1$\,mT due to the mixing of Zeeman sublevels by electric fields and strain internal to the diamond crystal. To overcome this limitation, one approach realized for a near-zero-field NV-ensemble sensor \cite{Zheng2019ZeroField} is to use, instead of an external bias field, the hyperfine magnetic field provided by the nitrogen nucleus of the NV center. For this to work, one needs a low-strain, narrow-line (e.g., $^{13}$C depleted) sample and polarization-selective microwave
excitation. In Ref.\,\cite{Zheng2019ZeroField}, circularly polarized microwaves were used to address specific hyperfine transitions in ODMR. The authors also describe how applying a modulated magnetic field allows one to perform zero-field magnetometry with linear polarized microwaves. Microwave-free zero-field ensemble magnetometry \cite{Ver2020_Weak_Mag_Field} and zero-field magnetometry with a single NV center \cite{Lenz2020magnetic} have also been demonstrated.

\subsection{Challenging spatial scales} 
\label{subsection:Challenging spatial scales}

Magnetic measurements are performed at spatial scales from atomic to astronomical, see Fig.\,\ref{Fig:MagnetomDifferentSpatialScales}. In this section, we discuss some of the extremes.

\subsubsection{Milli, micro, and nanoscales}
\label{subsubsection:small}

The size of a magnetometer limits both its spatial resolution and its integration into physical systems. When it comes to compact sensors, optically-pumped magnetometers stand out. First, the sensing element can be small or even nanoscopic; NV emission has been observed in single-digit nanometer diameter nanoparticles produced by milling~\cite{tisler2009fsp}, detonation~\cite{smith2009fnd, bradac2010ocb}, and oxidative etching~\cite{gaebel2012srn}. Second, optical and RF control eliminates the need for local electrical control. 

\begin{figure*}[t]
\centering
\includegraphics[width=\textwidth]{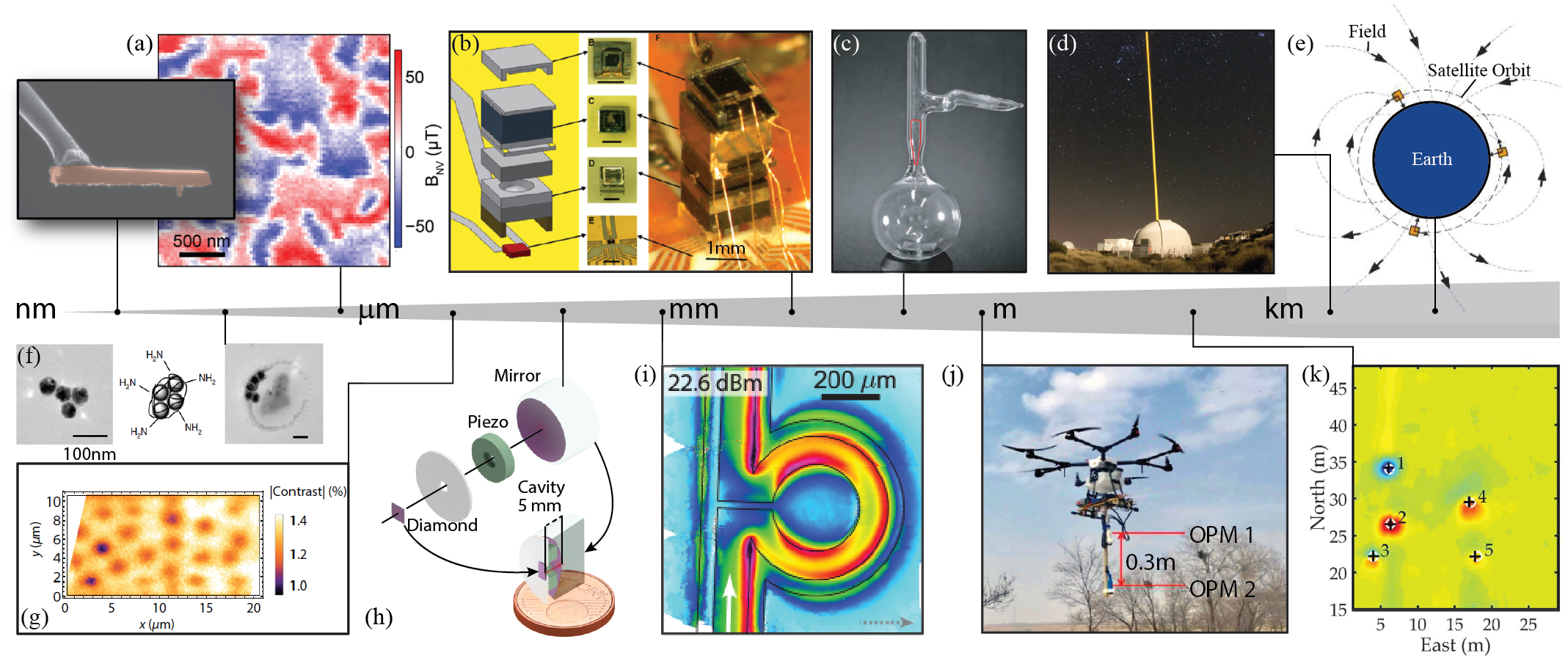}
\caption{Magnetometry at different length scales spanning 16 orders of magnitude. (a) diamond scanning-probe tip, antiferromagnetic domains imaged with a scanning single-NV probe \cite{Appel2019ScanningNV};  (b) chip-scale atomic magnetometer from~\cite{kitching2018csa};  (c) a typical-size vapor cell \cite{Balabas2010Minute}; (d) telescope launching a laser guide star for magnetometry with sodium in the mesosphere \cite{pedrerosbustos2018rsg}; (e) satellite-based magnetometry \cite{Zhou2018Satellite}; (f) nanodiamond hybrid sensor from \cite{zhang2018hnq}; (g) magnetic images of superconductor vortices obtained with NV centers \cite{schlussel2018wfi}; (h) 100\,$\mu$m NV ensemble sensor from \cite{chatzidrosos2017mce}. (i) Wide-field imaging of oscillating magnetic fields in the GHz range with NV centers \cite{Horsley2018Microwave}; (j) unmanned aerial vehicle fitted with vapor OPMs for magnetic surveying from \cite{Mu2020Drone}; (k) magnetic anomalies detected with the devices in (j).}
\label{Fig:MagnetomDifferentSpatialScales}
\end{figure*}

\textit{Nanoscale:} ODMR is observed in diamond particles of several nanometers in size~\cite{tisler2009fsp}, however practical magnetometry with such diamond remains challenging due to the high NV density required to ensure there is a color center in such a small nanoparticle~\cite{bradac2013enh}. Another issue is degradation in both photostability~\cite{bradac2013enh,kaviani2014pst} and coherence due to paramagnetic surface states~\cite{santawesin2019ods,myers2014psn}. Promisingly, advances towards addressing these challenges continue, including increased NV incorporation~\cite{baranov2011ehc}, advanced surface functionalization~\cite{sotoma2018eod}, and top-down fabrication of larger nanoparticles from bulk diamond~\cite{trusheim2014sfh}. Nanoparticle NV-magnetometers are particularly suitable when the nanoparticle sensor is functionalized to the sample (Fig.~\ref{Fig:MagnetomDifferentSpatialScales}f). For example, composite diamond and magnetic nanoparticle sensors can be utilized for enhanced temperature sensing in hybrid sensors discussed in Sec.\,\ref{Subsec:Hybrid}.

Nanoparticles were also utilized in the first nanoscale NV scanning probe experiments~\cite{maze2008nms,balasubramanian2008nim}, however most scanning probes today utilize nanoscale diamond tips fabricated from bulk diamond (Fig.\,\ref{Fig:MagnetomDifferentSpatialScales}a). High spatial resolution is enabled by the creation of NV centers nanometers from the single-crystal-diamond surface~\cite{ohashi2013ncn,ofori-okai2012spv} followed by reactive ion etching of the diamond~\cite{maletinsky2012rsd} to define the tip. Several commercial companies now supply these tips for scanning probe systems (e.g. Qnami AG, QZabre-LLC). State-of-the-art spatial resolution is typically in the 25-50\,nm range \cite{dovzhenko2018mtr,gross2017rsi, appel2015nmi,chang2017nic} and is primarily limited by the defect-surface distance. A new planar scanning microscopy modality was recently introduced \cite{Ernst2019PSM}, in which there is no tip at all, and a macroscopic diamond plate containing a shallow NV is scanned parallel to its surface.  NV scanning probe sensors were used for high-resolution imaging of a multitude of phenomena ranging from antiferromagnetic domains \cite{gross2017rsi} to skyrmion bubbles \cite{jenkins2019sss}

Beyond scanning probe applications, nanostructured diamond has additional features including an increased surface-to-volume ratio that has been leveraged to enhance solution NMR sensitivity~\cite{kehayias2017snm} and light-guiding capabilities for enhanced photon collection~\cite{momenzadeh2015ndw,mccloskey2020ewq}. Similar to nanoparticle diamond, in nanostructured diamond, charge stability, photostability, and spin coherence times are still outstanding challenges, however the single-crystal surface and control over the NV placement have enabled further studies towards addressing these problems~\cite{myers2017dqs, kim2015dns,santawesin2019ods}. 

Close proximity between sample and sensor can also be achieved utilizing ensembles of NV centers synthesized in a planar sheet at the diamond sample surface. This is for example used in Ref.~\cite{Horsley2018Microwave} to image oscillating magnetic fields in the microwave regime with high spatial resolution (Fig.\,\ref{Fig:MagnetomDifferentSpatialScales}i). 
Vapor-cell magnetometers are normally much larger but can also be produced to exhibit at least one dimension down to 40-1000\,nm as in the case of nanocells \cite{klinger2020pfn}. The reduced sensitivity to spatial scales is offset by the better signal-to-noise ratio of vapor-cell magnetometers, so they can still be used to detect, for example, sub-picomolar concentration of magnetic nanoparticles \cite{Bougas2018Picomolar}. In diamond, planar sheets of NV centers have enable wide-field imaging magnetic phenomena such as eddy currents~\cite{Chatzidrosos2019EddyNV}, magnetotactic bacteria~\cite{lesage2013omi}, and fluxtubes on a superconductor with a wide-field setup \cite{schlussel2018wfi} (Fig.\,\ref{Fig:MagnetomDifferentSpatialScales}g). If the close environment of the sensor is (super)-conductive, eddy currents in the material attenuating the delivery of oscillating magnetic fields reduce the sensor's performance. Mitigation strategies include all-optical and microwave-free sensing protocols discussed in sections \ref{subsec:all-optical}, \ref{subsec:MW-free}. The sensitivity to the electromagnetic properties in the sensor's local environment can and has been turned into an advantage to perform material-characterization studies with both vapor cell sensors \cite{Wickenbrock2014MIT, Wickenbrock2016eddy, Marmugi2019induction,Jensen2019eddy,Deans2020eddy} and diamond based sensors \cite{Chatzidrosos2019EddyNV}. The competitive edge of optically pumped magnetometers with respect to inductive eddy-current detection for nondestructive evaluation is the high sensitivity at low frequencies (high conductivity) and the high spatial resolution given by the potentially small sensor size and near-field measurements.

\textit{Micro to milliscale:} The above materials challenges in nanoscale NV magnetometry are less of a problem for micron-scale and larger devices. And for millimeter-scale applications, chip-scale atomic vapor magnetometers are now possible due to recent advancements in silicon chip-based atomic vapor cells~\cite{kitching2018csa} (Fig.\,\ref{Fig:MagnetomDifferentSpatialScales}b). The integration of micro-to-mm sensors and sensor arrays~\cite{IJsselsteijn2012} into biomedical devices is expected to impact an array of biomedical applications as discussed further in Sec.\,\ref{sec:bio}. Compact weight and size also is advantageous for satellite deployment (Sec.~\,\ref{subsubsection:space_and_remote}).

\subsubsection{Space-borne and remote magnetometry}
\label{subsubsection:space_and_remote}

At the opposite end of the size continuum is space-based magnetic sensing~\cite{Patton2013space}. Magnetometry can provide powerful information on the interactions between solar wind and solar system bodies and can further our understanding of planetary interiors. There are two main approaches: bring earth-made sensors into space (see e.g. Fig.\,\ref{Fig:MagnetomDifferentSpatialScales}e) or utilize atomic vapors already present near a heavenly body of interest. 

Atomic-vapor magnetometers have been employed in spacecraft missions since the 1961 Explorer X satellite. An overview of NASA's missions employing atomic vapor magnetometers is found in Ref.~\cite{Patton2013space}. The key considerations for satellite deployment include weight, volume (Sec.\,\ref{subsubsection:small}), cost, sensitivity and resilience to temperature changes (Sec.\,\ref{subsection:temperature}) and radiation (Sec.\,\ref{subsection:radiation}, as well as power consumption. Atomic magnetometers readily satisfy the dynamic range requirements from nT to tens of \textmu T for interplanetary and planetary exploration. The principal advantage of atomic magnetometers over the more compact and low-cost fluxgate vector magnetometers is the absolute nature of atomic sensors which do not require calibration. This property has led to a proposal to utilize atomic magnetometers in the evolved Laser Interferometer Space Antenna (eLISA) designed to test low-frequency gravitational wave radiation~\cite{Mateos2015nca}. eLISA requires low-frequency, high-sensitivity magnetic sensors because of the magnetic nature of the test masses in eLISA; a significant source of expected acceleration noise will be magnetic noise. 

Due to the compact size, broad temperature operation range, radiation resistance and ability for photo-electric readout, diamond-based OPMs are also taking their first steps into space~\cite{Hruby2018NVspace}. They may eventually compete with various solid-state flux-gate \cite{Miles2016Fluxgate} and magneto-resistance (impedance) \cite{diaz2009small} sensors whose drawback is that they are generally non-absolute and require calibration.  

Remarkably, it is even possible to detect magnetic fields in near space via the gas already present. For example, meteors deliver elements to the upper atmosphere, and there is a layer of atomic sodium of several km in thickness that exists at a height of 90-100 km above the Earth surface. These atoms can be excited with a beam of resonant light from a laser on the surface, creating a secondary (fluorescent) light source called a laser guide star (LGS). LGS systems using high-power lasers were developed for adaptive optics employed for real-time compensation of atmospheric perturbations and enabling near-diffraction-limited angular resolution for surface based telescopes with diameters up to tens of meters. For LGS based magnetometry, the laser light is amplitude, polarization, or frequency modulated in resonance with the Larmor frequency of the sodium atoms in the mesosphere, and the resonance is detected via a change in the intensity of laser-induced fluorescence~\cite{Higbie2011mms}.  Magnetometry with mesospheric sodium is an example of ``remote magnetometry'', to highlight the remote detection scheme of the highly local sensor. Several years after the original 2011 proposal~\cite{Higbie2011mms}, LGS magnetometry has been experimentally demonstrated ~\cite{kane2018lrm,pedrerosbustos2018rsg,fan2019rmm} with a current sensitivity of 30\,nT/$\sqrt{\text{Hz}}$~\cite{pedrerosbustos2018rsg} (Fig.\,\ref{Fig:MagnetomDifferentSpatialScales}d). Long-term applications include mapping the Earth's lithospheric magnetic fields, monitoring magnetic disturbances, long-term measurements on ionospheric currents and satellite-based explorations of other planets/moons with sodium atmospheres. 

There is another version of remote or ``stand-off'' magnetometry, where most of the magnetometer apparatus is at at a location spatially removed from the location where the field is measured (as in LGS magnetometry), except for just the sensor element (e.g., an alkali-vapor cell) and a retro-reflector for the laser laser light (see, for example, Ref.\,\cite{Patton2012remotely}). Such schemes, including LGS magnetometry, are examples of all-optical magnetometers discussed in Sec.\,\ref{subsec:all-optical}.

An interesting sensing application is that of remote-detection NMR \cite{mcdonnell2005nmr, xu2006mri,ledbetter2008zfr}. Here, the three NMR steps, polarization of the nuclei, encoding of the spatial or spin-composition information, and signal readout can all be separated in space. This could be useful, for instance, for imaging water flow inside a network of metal channels or a porous metal \cite{Xu2008PorousMetal}. Here, water is prepolarized in a strong magnetic field, then flows into the object to be imaged, where the spatial information is encoded into the time dependence of the polarization of the flowing sample, which is finally detected with a magnetometer downstream. Remote-detection NMR be useful for imaging large objects or when one cannot get the sensor into the object. The trade-off is that one cannot use the more common and time-efficient frequency-encoded NMR in this case and is reduced to using phase encoded, which generally leads to much longer image-acquisition times.

\subsection{Moving Platforms}
\label{sec:movingplatforms}

\begin{figure}[H]
\centering
\includegraphics[width=102mm]{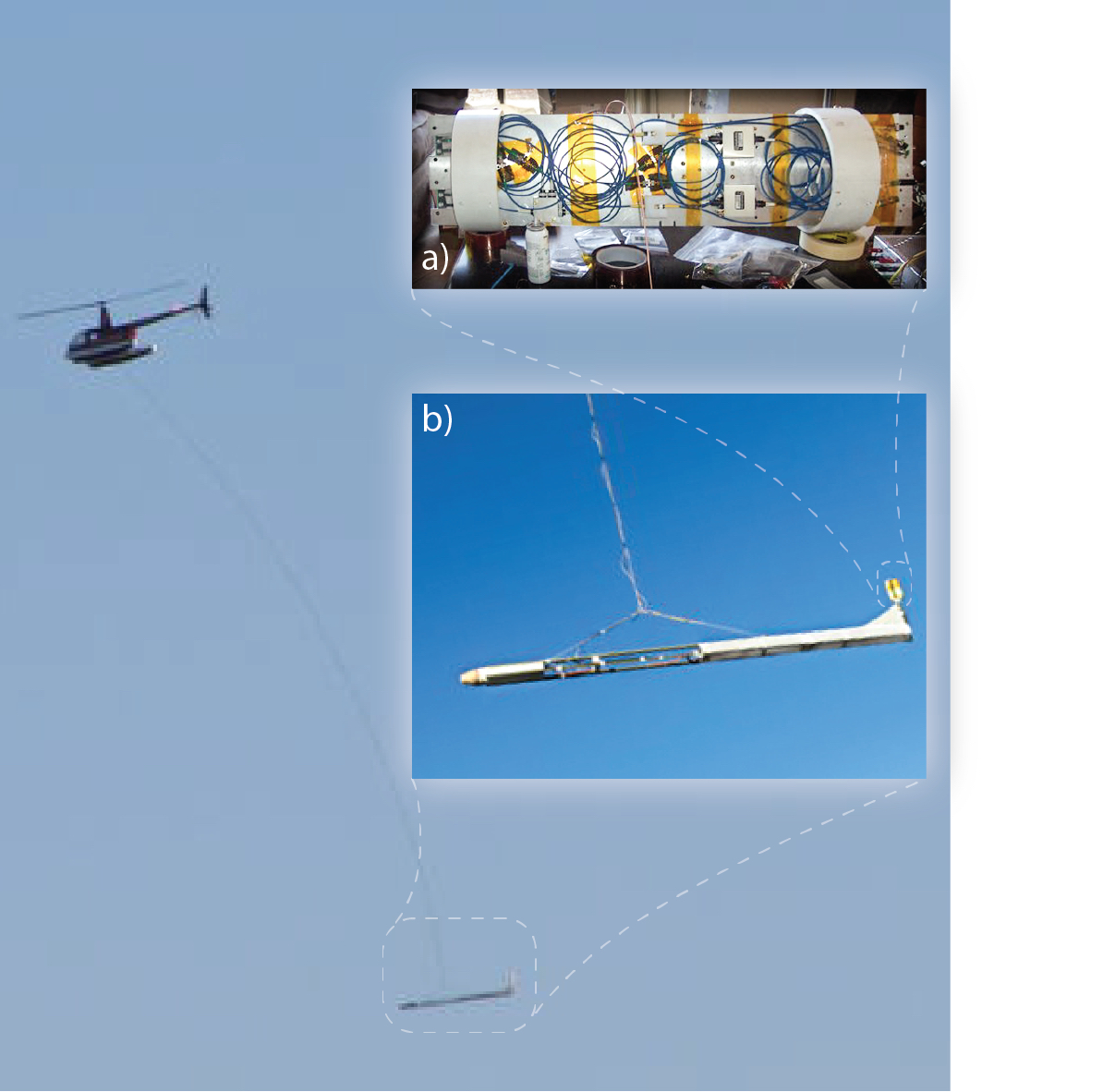}
\caption{Flight test, in 2012, of a towed magnetic gradiometer prototype developed by Southwest Sciences, Inc. in collaboration with UC Berkeley. Insets: a) gradiometer optical head assembly; b) aerodynamic ``bird'' with the gradiometer head mounted in the tail.}
\label{fig:Helicopter}
\end{figure}

As OPMs become more compact, there emerge applications on moving platforms, including handheld units, drilling assemblies~\cite{gooneratne2017dam}, unmanned aerial and underwater vehicles (UAV, UUV), satellites (Sec.\;\ref{subsubsection:space_and_remote}), cars and trucks, submarines and aircraft (Fig.\,\ref{fig:Helicopter}). Moving platforms pose challenges due to ``platform fields,'' vibrations, accelerations, and ``heading errors'' (see below).

When a magnetometer is operated in a passive mode, its optimal operation requires careful adjustment of the parameters, especially the ``working point’’ of the magnetic resonance. There are several strategies that can be combined to minimize the effect of vibrations on optimal operation including vibration isolation, monolithic integration~\cite{kitching2018csa,chatzidrosos2017mce,ibrahim2018rtq} (Fig.\,\ref{Fig:MagnetomDifferentSpatialScales}h), filtering (Sec.~\;\ref{subsection:mod_and_filtering}), closed-loop frequency-locking techniques \cite{Clevenson2018_Robust}, and device designs that minimize/eliminate permanent or induced magnetic fields from the sensor body. In the worst-case scenario, the sensor may completely lose the resonant signal. Loss of resonance is of particular concern for high-sensitivity, low bandwidth sensors. In such a situation, a self-oscillating device, see, for example, Ref.\,\cite{Higbie2006robust} and references therein, can be of practical advantage as it automatically restarts the optimum operation following a strong perturbation.
Measurement errors which arise due to the relative orientation of a magnetometer and the magnetic field, typically the Earth's magnetic field, are termed ``heading errors''~\cite{Budker2013ombook,oelsner2019she}. In OPMs, heading errors arise because of nonlinear Zeeman splitting (due to mixing of different hyperfine states by the magnetic field) and the light shift (due to mixing of the excited and ground states by the optical field). One practical but restrictive solution is to fix the sensor orientation with respect to the Earth's magnetic field~\cite{stuart1972efg,nabighian2010gem}. Modern solutions for cancelling the nonlinear Zeeman effect include utilizing double-modulated synchronous optical pumping~\cite{seltzer2007sop}, using high-order polarization moments~\cite{acosta2006nmo} free of nonlinearities, compensation with tensor light shifts \cite{jensen2009cnz}, spin-locking~\cite{bao2018snz}, and combining clockwise and counterclockwise circular polarization~\cite{oelsner2019she,ben-kish2010dzf}. 

Magnetic-anomaly detection (MAD) is often conducted with magnetometers on board aircraft or towed with helicopters (Fig.\,\ref{fig:Helicopter}). However, in recent years these platforms have been increasingly superseded with UAVs (Fig.\,\ref{Fig:MagnetomDifferentSpatialScales}j,k) and drones (see, for example, \url{https://areai.com/altius-500-4/}). In such systems, magnetic sensors have to withstand accelerations up 100\,g at launch, a challenge met by some of the modern vapor-cell based OPM such as Twinleaf's microSAM-2 total field magnetometer with no dead zones (\url{https://twinleaf.com/scalar/microSAM/}).   

Needless to say, OPMs do not hold a monopoly on operating on moving platforms. For example, an underwater SQUID based tensor gradiometer was recently deployed for scanning sea floor \cite{Chwala2019_Underwater}.  

\subsection{Radiation}
\label{subsection:radiation}
Sometimes, magnetic measurements need to be performed in environments with high levels of radiation, for example, in space, in the vicinity of or within nuclear reactors~\cite{landry1982smq}, or at the sites of nuclear accidents such as the one that occurred in 1986 in Chernobyl, Ukraine. Other examples of high-radiation environments are the beam tunnels and interaction halls at particle accelerators (see Sec.\,\ref{subsection:accelerator_facilities}), as well as fusion reactors~\cite{bolshakova2009phs}. 

\begin{figure}[t]
\centering
\includegraphics[width=2.5 in]{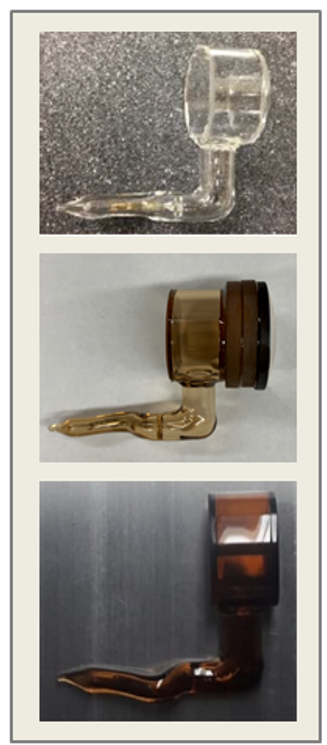}
\caption{The effect of radiation on magnetometer vapor cells \cite{Hovde2019rep}: (top) before irradiation, (middle) after 3.6\,Mrad (a typical dose for a mission to the Jovian moon Europa \cite{Buffington2014Europa}), (bottom) after 360\,Mrad. The cell in the middle has optics glued to it.}
\label{fig:IrradiatedCells}
\end{figure}

Because materials can be strongly affected by radiation, operation in high-radiation environments requires a particularly careful choice of materials and components used in a magnetometer and, ideally, extensive pre-operation testing with the type, energy, and radiation dose corresponding to a particular application. For example, vapor cells, coatings, and optical fibers should not undergo radiation-induced darkening (see Fig.\,\ref{fig:IrradiatedCells}), electronic components should maintain their integrity.
Fortunately, thanks to the considerable body of materials research in the context of space exploration, nuclear energy, and accelerator instrumentation, there exist databases of radiation resistant materials and components (for example, \cite{RadHardness}).

Diamond based magnetometers are of particular interest for sensing in high-radiation environments because diamond is among the best radiation-resistant materials. It is for this reason that diamond detectors are widely used in high-energy physics (see, for example, \cite{Bani2018dd} and references therein). 

\subsection{Magnetometry at accelerator facilities}
\label{subsection:accelerator_facilities}
Magnetic metrology is essential for the design and operation of accelerators and particle detectors \cite{Henrichsen2002magnetic_mapping} and accelerator based experiments often require precise and accurate magnetic-field measurements. An extreme example of this is the ``g-2'' experiment at Fermilab that aims to precisely measure the anomalous magnetic moment of the muon.
The magnetic-measurement system in this experiment \cite{Corrodi2020dap} employs 378 probes in fixed locations around a storage ring to constantly monitor magnetic-field drifts. These NMR sensors need to be periodically calibrated, which is done with a trolley which goes around the 45\,m long circumference and takes data every 2 to 3 days. The trolley-sensors themselves gets calibrated with precision absolute probes. The level of absolute calibration using $^3$He NMR has reached a remarkable level of 32\,ppb \cite{Farooq2020_3He_NMR}.

With all the remarkable high-precision results coming from advanced accelerator laboratories, one may not immediately realize that magnetometry in such an environment may, indeed, be especially challenging. As is well documented, for example, for the CERN Antiproton Decelerator hall \cite{Devlin2019sss} and the neutron electric-dipole moment (nEDM) experiment at the Paul Scherrer Institute \cite{Afach2014dsm}, there are multiple sources of spurious magnetic fields with diverse signatures such as pulsed extraction magnets, overhead cranes, and various other activities and devices associated with operation of the facility. Here active and passive shielding, as well as monitoring stray fields outside the experimental region are among the possible mitigation approaches. In the worst case scenario, the measurement needs to be temporarily interrupted while the perturbation is particularly intense \cite{Afach2014dsm}.

Apart from magnetic noise, other challenges include radiation (Sec.\,\ref{subsection:radiation}) and poor climate control in large experimental halls.

\subsection{Magnetometry in urban environment}
\label{subsection:urban_magnetometry}
Cities with their numerous and diverse sources of magnetic noise associated with power lines, surface and underground rail systems, automobile traffic, operating machinery and buildings, are particularly challenging for magnetometry. On the flip side, performing magnetic measurements in the cities can provide unique information on urban patterns whose analysis could be useful for the understanding and optimization of energy utilization, environmental monitoring, or for threat detection via detecting of unusual magnetic activity due to structural changes or clandestine activity.

In the recent work  of \cite{Bowen2019}, a network of global positioning system (GPS) synchronized flux-gate magnetometers with a 100\,pT/$\sqrt{\text{Hz}}$ sensitivity sampled at $\approx$4\,kHz was used to study the magnetic environment in and around the city of Berkeley, CA, USA. It was found (see also the references within \cite{Bowen2019}) that the dominant magnetic-noise sources were associated with the Bay Area Rapid Transit (BART) system, and ways to identify much weaker signals in the presence of this dominant signal based on wavelet analysis applied to the network data were discussed. This work is the first step towards identifying the ``magnetic profile'' of different cities.

\section{Biosensitive measurements}
\label{sec:bio}

\subsection{Magnetometry at the cellular level}
\label{subsec:magnetometry_at_cellular_level}

Significant interest has arisen in the use of magnetometry, and specifically NV centers in diamond, to study biological processes and biomolecular structure of living organisms. Many of the challenges for noninvasive sensing in these systems arise from the scale of the sensor and apparatus, relative to the probed system, which is often at the cellular level. The relative simplicity of NV setups, as compared to the large cryogenic apparatus required for SQUID systems, has made them attractive for use in detection of biological signals, for example, detecting action potentials in living systems, or probing bacterial cells placed on a diamond surface with NVs implanted just below. Such a setup allows for not only sensing, but optical imaging of the magnetic field, providing detailed information about production of magnetosomes in bacterial cultures or in localizing magnetic markers or nanoparticle tracers \cite{lesage2013omi,kuwahata2020}. Similar setups have been used to perform magnetic resonance imaging of spin-labelled proteins in cell membranes \cite{steinert2013}. In Ref.\,\cite{barry2016}, it was shown that sensors could be placed as close as 10\,$\mu$m from the surface of an undissected marine fanworm Myxicola infundibulum. At this distance, the magnetic signal associated with stimulated action potentials were as large as 1\,nT, compared to the 15\,pT/$\sqrt{\textrm{Hz}}$ sensitivity of the sensor, demonstrating the unique combination of sample proximity and high sensitivity afforded by NV centers in diamond for biological applications. 

The low cytotoxicity of diamond and robust fluorescence properties have motivated the use of nanodiamonds with NV centers as exquisitely sensitive, local probes of the fields even within these living systems, where the challenges for magnetometry with SQUIDs, fluxgate sensors, or atomic vapor cells could be insurmountable \cite{balasubramanian2014, van2018nanodiamonds, chipaux2018nanodiamonds}. The magnetic properties of the NV spin structure enable magneto-optical nanoscopy, wherein fluorescent nanodiamonds can be utilized as photostable tracers in biological systems, tracking intercellular protein transport \cite{kuo2013fluorescent}, or as magnetic tracers localized in a magnetic field gradient, in analogy to MRI technique \cite{lin2019fluorescent}.

Significant progress has been made in mitigating the challenges associated with nanodiamond based magnetometry (as opposed to bulk), including random orientations of the NV centers and broad thermal distributions of nanodiamond ensembles \cite{wojciechowski2019,wojciechowski2018precision}.  Opening up further applications, fluorescent nanodiamonds were recently combined with biomimetic substrates used in the laboratory. ODMR spectra were obtained from live differentiated neural stem cells functioning as a connected neural network grown on fluorescent nanodiamond embedded nanofibers \cite{price2019}. See \cite{chipaux2018nanodiamonds} and references therein for a detailed review on the use of nanodiamonds in cells.
 
Vapor cell OPMs have generally been considered too large in size to have the resolution necessary for magnetic detection in cellular biology. Recent advances in chip-scale vapor cell fabrication have made possible cm-size OPMs based on mm-scale cells (Sec.\,\ref{sec:op_mags}), but below these scales, spin-destructive collisions with the cell walls dominate the spin relaxation rates, creating a trade-off between vapor cell size and sensitivity \cite{Kim2016, allred2002high}. One approach around the larger vapor-cell size is the use of flux guides, as described in  Sec.\,\ref{subsection:Flux_engineering}.

\subsection{Inside the human body}
\label{subsec:inside_human_body}

Virtually all processes within the human body are controlled and regulated via electrochemical signals. Non-contact measurements of magnetic fields associated with nerve and muscle activity are of great interest because these signals can communicate detailed temporal and spatial information \cite{williamson2013}. This information is highly complementary to existing techniques to measure electromagnetic activity within the human body, such as electroencephalography (EEG) or electromyography (EMG), as well as to those techniques measuring the dynamics of blood flow (hemodynamics) associated with neural activity, such as functional magnetic resonance imaging (fMRI), positron-emission tomography (PET), functional near-infrared spectroscopy (fNIRS), and single-photon emission computed tomography (SPECT). From a medical perspective, magnetic fields offer an additional modality by which to study the muscle and nervous system of the body that is distinct from electric sensing techniques and magnetic resonance imaging \cite{williamson1981,romani1984}.

Whether originating from nerves or muscles, the detected magnetic signal is the sum of fields from many individual cells firing respective action potentials. The aggregate detected signal, nevertheless, remains within the reach of only the most sensitive of magnetometers \cite{schenck1996,williamson2013}. The highest-amplitude magnetic signals from the body emanate from the heart, with magnetic fields of up to 100\,pT above the skin \cite{koch2004}. Meanwhile, fields due to neural activity deep within the temporal lobe of the brain are typically below the 10\,fT level at the scalp, and require a high number of sensors to accurately reconstruct the source location \cite{vrba2002}. 

\subsection{Magnetometry of the brain, nerves and  muscles}
\label{subsec:brain_magnetometry}

Despite the challenges, magnetometry of brain function, or magnetoencephalography (MEG), is a powerful tool with high spatial and temporal resolution, providing a direct measure of the electrical activity of neurons, in contrast to other imaging techniques such as fMRI, PET or EEG. Furthermore, MEG is completely noninvasive, making it an attractive option for both medical and technological applications such as brain-computer interfaces (BCI) \cite{mellinger2007meg}.

Clinically relevant measurements of magnetic fields from brain activity require both high sensitivity to detect the fields and high sensor density to perform accurate source localization \cite{riaz2017}. Until recently, only SQUID technologies were mature enough to meet these requirements, generating a global MEG industry. SQUID MEG systems mitigate the environmental challenges of detecting brain activity in a hospital by operating in magnetically shielded environments and using gradiometric sensors. Typically, these devices incorporate hundreds of sensors into a rigid head enclosure that is connected to a dewar that provides the necessary cryogenic cooling. While the rigidity and fixed position of the enclosure is ideal for magnetometer operation and source localization, incorporating sensitive magnetometry into natural environments for subjects, ideally without magnetic shielding, remains an open challenge.

Widely considered, therefore, are OPMs and NV magnetometers, which, due to their small size, can move freely with the user, untethered by cryogenic cooling.  While NV magnetometers are rapidly approaching sensitivities necessary to measure magnetic fields from the brain\,\cite{barry2016}, OPMs have already overcome this threshold. Recent demonstrations with OPMs have shown that they can be placed closer to the scalp than comparable SQUID sensors, where the benefit is not only in yielding higher signal amplitudes from neurological magnetic sources, but also in improved spatial information density, which outweighs benefits of higher sensitivity or of more sensors \cite{boto2016, riaz2017, iivanainen2017, iivanainen2019}. To this end, multichannel magnetometry with OPMs within millimeters of the scalp has been studied with up to 25 magnetometers \cite{borna2017, alem2017}, including demonstrations of wearable systems\,\cite{boto2018}.

Yet movement of these sensors places stringent requirements on the background field. A subject may move their head within a 30\,cm cubed region, which in turn must be homogeneous to a high degree. Gradients as small as 1\,nT/cm could cause large artifacts that can either swamp or be confused with brain signals. As a result, sensor tracking (for example optically, as mentioned in Sec.\,\ref{Subsec:Hybrid}) within well characterized field regions, and active shielding where the background-field noise is actively zeroed using the magnetometer signal \cite{zhang2020recording,zhang2020active}, can offer benefits over passive shielding (Sec.\,\ref{subsec:Shielding}).

An important challenge in MEG is in source localization, which requires solving an ill-posed inverse problem. Well-defined head shape (using, for example, MRI of the subject) can constrain the solutions, but also crucial is sufficiently dense field mapping with vector information. OPMs and NVs can be readily used as vector sensors in at least two dimensions, and NVs are particularly well suited to high densities of individual magnetometers. While OPMs are limited by vapor-cell size, relatively high sensor densities can still be achieved. At these sensor densities, however, cross-talk between sensors and miniaturization with heating could be problematic. Cross-talk can be addressed via prior characterization, compensation and shielding, or all-optical implementations (Sec\,\ref{subsec:all-optical}).

Magnetic detection of neural and muscle activity of other body parts is also of clinical interest. While facing similar implementation challenges as magnetoencephalography, magnetic signals from muscles are on the order of 1\,pT near the surface of the skin. Measurements of muscle signals, or magnetomyography, have been performed with the sensors described above, and, in combination with other technologies such as TMS, can offer tools to study the human motor system  with high spatio-temporal resolution (see Sec.\,\ref{subsec:Shielding}). While these relatively large muscle signals are useful when the object of interest is the muscle itself, they can obscure magnetic signals that accompany nerve transmission, which are also of medical interest~\cite{elzenheimer2018mng}. Internal magnetometry with, for example, magnetic endoscopes, offer higher SNR (due to the reduced distance to sources) and spatial filtering, with added challenges of biocompatibility and further miniaturization. 
Needle-sized magnetometers, previously implemented with giant magnetoresistance sensors~\cite{shirzadfar2015ntg}, are possible with diamond OPMs and could measure previously inaccessible in-vivo fields. This could enable new diagnostic tools in several fields, for example, cardiology, functional neuro-imaging and brain surgery. In the brain (or in the peripheral nerves or the spine), magnetometers could measure neural activity with sub-mm resolution, which would allow, given the appropriate sensitivity, functional monitoring of individual nerves or structures. Additionally, detection of oscillating magnetic fields can be used to discriminate different tissue types in the context of conductivity measurements or low-field NMR sensing. In particular, discrimination between healthy and tumorous tissue using local conductivity measurement~\cite{Foster1981Cancer1, Haemmerich2009Cancer2}, appears promising and might be within reach of optimized laboratory OPMs~\cite{Chatzidrosos2019EddyNV}. In the field of functional neuroimaging via MEG (cf. \ref{subsec:brain_magnetometry}), optically pumped alkali vapor-cell magnetometers are already employed in new modalities due to their reduced sensor-to-skull distance and small overall sensor size. Going a step further and providing magnetic information from inside the body using needle magnetometers could dramatically improve source location, especially in conjunction with a dense sensor network around the skull.

\subsection{Plant magnetometry}
\label{subsec:plant_magnetometry}

Members of the plant kingdom have been probed in exquisite detail using a variety of sensing modalities. Yet literature covering magnetic sensing of plant processes is limited \cite{corsini2011search}, perhaps owing to a) the extremely small expected signal size, b) difficulties in developing reliable stimulation protocols in the plant, and c) the relatively long timescale of plant signalling (1-5\,s) - a regime of abundant low-frequency noise. Since the plant action potential amplitudes are between 5-10 times smaller than in animals and the propagation speeds are typically four orders of magnitude smaller \cite{szechynska2017electrical}, there are grave implications for the accompanying magnetic field amplitudes generated by most plants, again strictly limiting the magnetometer types that could be used for detecting. 

Nevertheless, successful attempts have been made, in particular with the use of SQUID magnetometers to measure signalling in wounded bean plants \cite{jazbinsek2000magnetic}, and OPMs to measure stimulated action potentials in Venus Flytraps~\cite{fabricant2020api}.

\section{Concluding remarks}
As illustrated in this review, OPM's have reached the operational sensitivity to enable a wide range of applications. 
It is the hope of  the authors that this review would provide the reader with a  set of  general and specific ideas of how to meet various  diverse  challenges arising in the way of real-life precision magnetic (or other) measurements. Extrapolating from the rapid progress made over the past decade, we can expect a future that includes not just higher sensitivity but further miniaturization (with compact shielding), resiliency to harsher environments, more extensive networks enabled by modern computational techniques, and further integration into multi-functional hybrid sensors, all of which will open the application space further. 

There are also many open questions. While this review mostly focuses on passive sensing devices, self-oscillating magnetometers \cite{Higbie2006robust}, lasers operating on magnetically sensitive transitions, and spin masers \cite{yoshimi2002nuclear,bevington2019alkali,jiang2019floquet} belong to the category of ``active'' sensing devices. The spectral width of the signal generated by active devices could be orders of magnitude narrower than the width of the underlying atomic line. It is thus tempting to ask: may such devices offer fundamental advantages over their ``passive'' counterparts? In all numerous specific cases known to us, the answer is no; however, we are not aware of a general proof that this should be the case. Even if active devices cannot fundamentally outperform their passive counterparts, they are already proven to often be advantageous in practice, for example with moving platforms (Sec.\,\ref{sec:movingplatforms}) and for detecting low-frequency magnetic fields. Additional open questions include the relative fundamental and practical performance of intrinsic vs. composite gradiometers (Sec.\,\ref{subsubsection:Differential_magnetometry}), whether solid-state OPMs will reach sensitivities comparable to those of vapor-cell OPMs, and if there are far superior solid state defects, relative to the NV center, for sensing applications~\cite{dreyer2018fpc,bassett2019qdd}.

Finally, this review would be incomplete without mentioning the high-precision Lorentz-invariance tests carried out with atomic sensors at the South Pole \cite{Smiciklas2014SouthPole}, as an example of a challenging measurement in more than one sense.

\section*{Acknowledgements}

The authors are grateful to L. Bougas, I. Fescenko, M. Jiang, D.\,C. Hovde, J. Hruby, P. Kehayias, K. Kirch, T. Kornack, P. Koss, R. Liu, P. Maletinsky, S. Prawer, M. Romalis, T. Scholtes, S. Ulmer, Y. Semertzidis, and R. Zhang for useful discussions. This work was supported by the EU FET-OPEN Flagship Project ASTERIQS (Action 820394) and the German Federal Ministry of Education and Research (BMBF) within the Quantumtechnologien program (FKZ 13N14439 and FKZ 13N15064). DB was supported in part by the DFG Project ID 390831469:  EXC 2118 (PRISMA+ Cluster of Excellence). DB also received support from the European Research Council (ERC) under the European Union Horizon 2020 Research and Innovation Program (grant agreement No. 695405) and from the DFG Reinhart Koselleck Project. KMCF recognizes support from the HIM Visiting Scientist Program and the National Science Foundation under Grant No. CHE-1607869.

\section*{Data Availability}
Data sharing is not applicable to this article as no new data were created or analyzed in this study.

\bibliography{Literature.bib}

\end{document}